\documentclass{article}

\usepackage[margin=1.5cm, includefoot, footskip=30pt]{geometry}
\usepackage{hyperref}
\usepackage{multicol}
\usepackage{booktabs}
\usepackage{xstring}
\usepackage{subcaption}
\usepackage{graphicx}
\usepackage{amsmath}
\usepackage{standalone}
\usepackage{fancyvrb}
\usepackage{authblk}

\usepackage{pdflscape}

\usepackage{tikz}
\usetikzlibrary{calc, shapes, patterns}

\title{Reinforcement Learning Produces Dominant Strategies for the
Iterated Prisoner's Dilemma}
\author[1]{Marc Harper}
\author[2]{Vincent Knight}
\author[3]{Martin Jones}
\author[4]{Georgios Koutsovoulos}
\author[2]{Nikoleta E. Glynatsi}
\author[3]{Owen Campbell}

\affil[1]{Google Inc., Mountain View, CA, USA}
\affil[2]{Cardiff University, School of Mathematics, UK}
\affil[3]{Not affiliated}
\affil[4]{INRA, Université Côte d‘Azur, CNRS, ISA, France}
\date{}

\begin{document}

\maketitle

\begin{abstract}
    We present tournament results and several powerful strategies for the Iterated
    Prisoner's Dilemma created using reinforcement learning techniques
    (evolutionary and particle swarm algorithms). These strategies are
    trained to perform well against a corpus of over 170 distinct
    opponents, including many well-known and classic strategies. All
    the trained strategies win standard tournaments against the total collection
    of other opponents. The trained strategies and one particular human made
    designed strategy are the top performers in noisy tournaments also.
\end{abstract}

\section{Introduction}\label{sec:introduction}

The Iterated Prisoner's Dilemma (IPD) is a common model in game theory,
frequently used  to understand the evolution of cooperative behaviour from complex
dynamics \cite{Axelrod1984}.

This manuscript uses the Axelrod library \cite{knight2016open, axelrodproject},
open source software for conducting IPD research with
reproducibility as a principal goal. Written in the Python programming language,
to date the library contains source code contributed by over 50 individuals
from a variety of geographic locations and technical backgrounds. The library
is supported by a comprehensive test suite that covers all the intended
behaviors of all of the strategies in the library, as well as the features that
conduct matches, tournaments, and population dynamics.

The library is continuously developed and as of version 3.0.0, the library
contains over 200 strategies, many from the scientific literature, including
classic strategies like Win Stay Lose Shift \cite{nowak1993strategy} and
previous tournament winners such as OmegaTFT \cite{slany2007some}, Adaptive
Pavlov \cite{li2007design}, and ZDGTFT2 \cite{Stewart2012}.

Since Robert Axelrod's seminal tournament \cite{Axelrod1980a}, a number of
IPD\@ tournaments have been undertaken and are summarised in
Table~\ref{tbl:tournaments}. Further to the work described in
\cite{knight2016open} a regular set of standard, noisy~\cite{Bendor1991} and
probabilistic ending~\cite{Axelrod1980b} tournaments are carried out as more
strategies are added to the Axelrod library.
Details and results are available here:
\url{http://axelrod-tournament.readthedocs.io}. This work presents a detailed
analysis of a tournament with 176 strategies (details given
in Section~\ref{sec:results}).

\begin{table}[!hbtp]
    \begin{center}
        \begin{tabular}{ccccc}
            \toprule
            Year     & Reference                  & Number of Strategies & Type     & Source Code\\
            \midrule
            1979     & \cite{Axelrod1980a}        & 13                   & Standard & Not immediately available\\
            1979     & \cite{Axelrod1980b}        & 64                   & Standard & Available in FORTRAN\\
            1991     & \cite{Bendor1991}          & 13                   & Noisy    & Not immediately available\\
            2002     & \cite{Stephens2002}        & 16                   & Wildlife & Not applicable\\
            2005     & \cite{kendall2007iterated} & 223                  & Varied   & Not available \\
            2012     & \cite{Stewart2012}         & 13                   & Standard & Not fully available \\
            2016     & \cite{knight2016open}       & 129                  & Standard & Fully available \\
            \bottomrule
        \end{tabular}
    \end{center}
    \caption{An overview of a selection of published tournaments. Not all
             tournaments were `standard' round robins; for more details
             see the indicated references.}\label{tbl:tournaments}
\end{table}

In this work we describe how collections of strategies in the Axelrod library
have been used to train new strategies specifically to win IPD tournaments.
These strategies are trained using generic strategy archetypes based on e.g.\
finite state
machines, arriving at particularly effective parameter choices through
evolutionary or particle swarm algorithms. There are several
previous publications that use evolutionary algorithms to
evolve IPD strategies in various circumstances
\cite{ashlock2006training, Ashlock2015, Ashlock2006,
      ashlock2014shaped, Ashlock2014, barlow2015varying,
      fogel1993evolving, marks1989niche, sudo2015effects,
      vassiliades2010multiagent}. See also \cite{Gaudesi2016} for a
strategy trained to win against a collection of well-known IPD opponents and see
\cite{franken2005particle} for a prior use of particle swarm algorithms. Our
results are unique in that we are able to train against a large and diverse
collection of strategies available from the scientific literature.
Crucially, the
software used in this work is openly available and can be used to train strategies
in the future in a reliable manner, with confidence that the opponent strategies
are correctly implemented, tested and documented.
Moreover, as of the time of writing, we claim that this work contains the best
performing strategies for the Iterated Prisoner's Dilemma.

\section{The Strategy Archetypes}

The Axelrod library now contains many parametrised strategies trained using
machine learning
methods. Most are deterministic, use many rounds of memory, and perform
extremely well in tournaments as will be discussed in Section~\ref{sec:results}.
Training of these strategies will be discussed in Section~\ref{sec:methods}.
These strategies can encode a variety
of other strategies, including classic strategies like Tit For Tat
\cite{Axelrod1980},
handshake strategies, and grudging strategies, that always defect after
an opponent defection.

\subsection{LookerUp}\label{sec:lookerup}

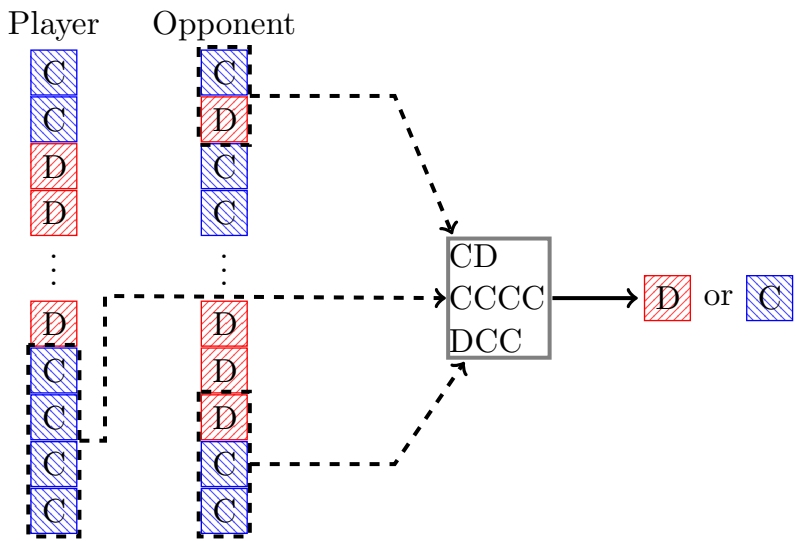
\begin{figure}[!hbtp]
    \centering
    \begin{tikzpicture}[scale=.9]
    \tikzstyle{cooperator} = [rectangle, pattern=north west lines, pattern
        color=blue!70, draw=blue, text width=.25cm, outer sep = 2pt]
    \tikzstyle{defector} = [rectangle, pattern=north east lines, pattern
        color=red!70, draw=red, text width=.25cm, outer sep = 2pt]

    \node (A) at (-.5, 0) {Player};
    \node (A1) at ($(A) + (0, -.55)$) [cooperator] {C};
    \node (A2) at ($(A1) + (0, -.55)$) [cooperator] {C};
    \node (A3) at ($(A2) + (0, -.55)$) [defector] {D};
    \node (A4) at ($(A3) + (0, -.55)$) [defector] {D};
    \node (A5) at ($(A4) + (0, -.55)$) {$\vdots$};
    \node (A6) at ($(A5) + (0, -.75)$) [defector] {D};
    \node (A7) at ($(A6) + (0, -.55)$) [cooperator] {C};
    \node (A8) at ($(A7) + (0, -.55)$) [cooperator] {C};
    \node (A9) at ($(A8) + (0, -.55)$) [cooperator] {C};
    \node (A10) at ($(A9) + (0, -.55)$) [cooperator] {C};

    \node (B) at ($(A) + (2, 0)$) {Opponent};
    \node (B1) at ($(B) + (0, -.55)$) [cooperator] {C};
    \node (B2) at ($(B1) + (0, -.55)$) [defector] {D};
    \node (B3) at ($(B2) + (0, -.55)$) [cooperator] {C};
    \node (B4) at ($(B3) + (0, -.55)$) [cooperator] {C};
    \node (B5) at ($(B4) + (0, -.55)$) {$\vdots$};
    \node (B6) at ($(B5) + (0, -.75)$) [defector] {D};
    \node (B7) at ($(B6) + (0, -.55)$) [defector] {D};
    \node (B8) at ($(B7) + (0, -.55)$) [defector] {D};
    \node (B9) at ($(B8) + (0, -.55)$) [cooperator] {C};
    \node (B10) at ($(B9) + (0, -.55)$) [cooperator] {C};

    \draw[very thick, black, dashed] ($(B1) + (-.3, .3)$)  rectangle ($(B2) + (.3, -.3)$);
    \draw[very thick, black, dashed] ($(A7) + (-.3, .3)$)  rectangle ($(A10) + (.3, -.3)$);
    \draw[very thick, black, dashed] ($(B8) + (-.3, .3)$)  rectangle ($(B10) + (.3, -.3)$);

    \node (opp_start) at (4, -2.7) [right] {CD};
    \node (player_history) at (4, -3.2) [right] {CCCC};
    \node (opp_history) at (4, -3.7) [right] {DCC};

    \draw [very thick, black, dashed, ->] ($(B1)!0.5!(B2) + (.3, 0)$) --
        ($(B1)!0.5!(B2) + (2, 0)$) -- ($(opp_start) + (-.25, .25)$);
    \draw [very thick, black, dashed, ->] ($(A7)!0.5!(A10) + (.3, 0)$) --
        ($(A7)!0.5!(A10) + (.6, 0)$) -- ($(A7)!0.5!(A10) + (.6, 1.7)$) --
        ($(player_history) + (-.6, 0)$);
    \draw [very thick, black, dashed, ->] ($(B8)!0.5!(B10) + (.3, 0)$) --
        ($(B8)!0.5!(B10) + (2, 0)$) -- ($(opp_history) + (-.25, -.25)$);


    \draw [very thick, gray] ($(opp_start) + (-.3, .2)$) rectangle
    ($(opp_history) + (.75, -.2)$);

    \node (output) at ($(player_history) + (2, 0)$) [defector] {D};
    \node (or) at ($(output) + (.6, 0)$) {or};
    \node at ($(or) + (.6, 0)$) [cooperator] {C};

    \draw [->, very thick] ($(player_history) + (.65, 0)$) -- (output);
\end{tikzpicture}
    \caption{Diagrammatic representation of the Looker up Archetype.}
    \label{fig:lookerup}
\end{figure}

The LookerUp strategy is based on a lookup table and encodes a set of
deterministic responses based on the opponent's first $n_1$ moves, the
opponent's last $m_1$ moves, and the players last $m_2$ moves. If $n_1 > 0$ then
the player has infinite memory depth, otherwise it has depth $\max(m_1, m_2)$.
This is illustrated diagrammatically in Figure~\ref{fig:lookerup}.

Training of this strategy corresponds to finding maps from partial histories to
actions, either a cooperation or a defection. Although various
combinations of $n_1, m_1,$ and $m_2$ have been tried, the best performance at
the time of
training was obtained for $n_1 = m_1 = m_2 = 2$ and generally for $n_1 > 0$.
A strategy
called EvolvedLookerUp2\_2\_2 is among the top strategies in the library.

This archetype can be used to train deterministic memory-$n$ strategies with the
parameters $n_1=0$ and $m_1=m_2=n$. For $n=1$, the resulting strategy cooperates
if the last round was mutual cooperation and defects otherwise, known as Grim or
Grudger.

Two strategies in the library, Winner12 and Winner21, from \cite{Mathieu2015},
are based on lookup tables for $n_1 = 0$, $m_1 = 1$, and $m_2=2$. The strategy
Winner12 emerged in less than 10 generations of training in our framework using
a score maximizing objective. Strategies nearly identical to Winner21 arise
from training with a Moran process objective.

\subsection{Gambler}\label{sec:gambler}

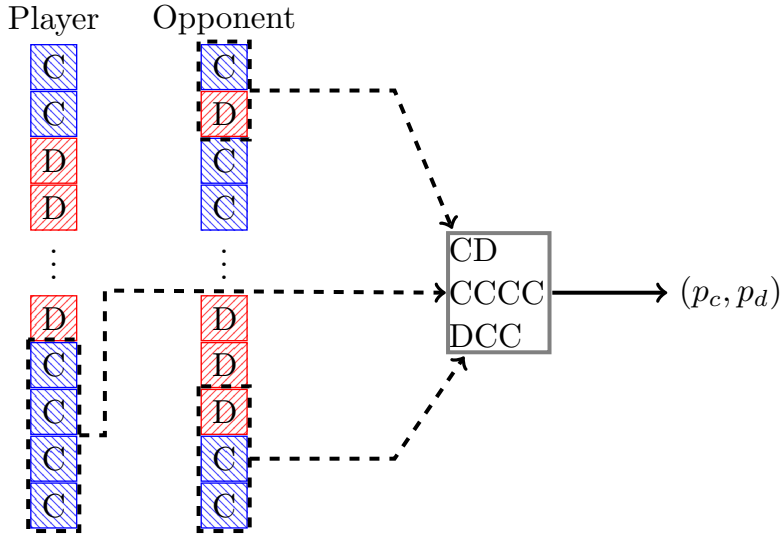
\begin{figure}[!hbtp]
    \centering
    \begin{tikzpicture}[scale=.9]
    \tikzstyle{cooperator} = [rectangle, pattern=north west lines, pattern
        color=blue!70, draw=blue, text width=.25cm]
    \tikzstyle{defector} = [rectangle, pattern=north east lines, pattern
        color=red!70, draw=red, text width=.25cm]

    \node (A) at (-.5, 0) {Player};
    \node (A1) at ($(A) + (0, -.55)$) [cooperator] {C};
    \node (A2) at ($(A1) + (0, -.55)$) [cooperator] {C};
    \node (A3) at ($(A2) + (0, -.55)$) [defector] {D};
    \node (A4) at ($(A3) + (0, -.55)$) [defector] {D};
    \node (A5) at ($(A4) + (0, -.55)$) {$\vdots$};
    \node (A6) at ($(A5) + (0, -.75)$) [defector] {D};
    \node (A7) at ($(A6) + (0, -.55)$) [cooperator] {C};
    \node (A8) at ($(A7) + (0, -.55)$) [cooperator] {C};
    \node (A9) at ($(A8) + (0, -.55)$) [cooperator] {C};
    \node (A10) at ($(A9) + (0, -.55)$) [cooperator] {C};

    \node (B) at ($(A) + (2, 0)$) {Opponent};
    \node (B1) at ($(B) + (0, -.55)$) [cooperator] {C};
    \node (B2) at ($(B1) + (0, -.55)$) [defector] {D};
    \node (B3) at ($(B2) + (0, -.55)$) [cooperator] {C};
    \node (B4) at ($(B3) + (0, -.55)$) [cooperator] {C};
    \node (B5) at ($(B4) + (0, -.55)$) {$\vdots$};
    \node (B6) at ($(B5) + (0, -.75)$) [defector] {D};
    \node (B7) at ($(B6) + (0, -.55)$) [defector] {D};
    \node (B8) at ($(B7) + (0, -.55)$) [defector] {D};
    \node (B9) at ($(B8) + (0, -.55)$) [cooperator] {C};
    \node (B10) at ($(B9) + (0, -.55)$) [cooperator] {C};

    \draw[very thick, black, dashed] ($(B1) + (-.3, .3)$)  rectangle ($(B2) + (.3, -.3)$);
    \draw[very thick, black, dashed] ($(A7) + (-.3, .3)$)  rectangle ($(A10) + (.3, -.3)$);
    \draw[very thick, black, dashed] ($(B8) + (-.3, .3)$)  rectangle ($(B10) + (.3, -.3)$);

    \node (opp_start) at (4, -2.7) [right] {CD};
    \node (player_history) at (4, -3.2) [right] {CCCC};
    \node (opp_history) at (4, -3.7) [right] {DCC};

    \draw [very thick, black, dashed, ->] ($(B1)!0.5!(B2) + (.3, 0)$) --
        ($(B1)!0.5!(B2) + (2, 0)$) -- ($(opp_start) + (-.25, .25)$);
    \draw [very thick, black, dashed, ->] ($(A7)!0.5!(A10) + (.3, 0)$) --
        ($(A7)!0.5!(A10) + (.6, 0)$) -- ($(A7)!0.5!(A10) + (.6, 1.7)$) --
        ($(player_history) + (-.6, 0)$);
    \draw [very thick, black, dashed, ->] ($(B8)!0.5!(B10) + (.3, 0)$) --
        ($(B8)!0.5!(B10) + (2, 0)$) -- ($(opp_history) + (-.25, -.25)$);


    \draw [very thick, gray] ($(opp_start) + (-.3, .2)$) rectangle
    ($(opp_history) + (.75, -.2)$);

    \node (output) at ($(player_history) + (2, 0)$) [right] {\((p_c, p_d)\)};

    \draw [->, very thick] ($(player_history) + (.65, 0)$) -- (output);
\end{tikzpicture}
    \caption{Diagrammatic representation of the Gambler Archetype.}
    \label{fig:gambler}
\end{figure}

Gambler is a stochastic variant of LookerUp. Instead of deterministically
encoded moves the lookup table emits probabilities which are
used to choose cooperation or defection.
This is illustrated diagrammatically in Figure~\ref{fig:gambler}.

Training of this strategy corresponds to finding maps from histories to
a probability of cooperation. The library includes a strategy trained
with $n_1 = m_1 = m_2 = 2$ that is \emph{mostly deterministic}, with 52 of the 64
probabilities being 0 or 1. At one time this strategy outperformed
EvolvedLookerUp2\_2\_2.

This strategy type can be used to train arbitrary memory-$n$ strategies. A
memory one strategy called PSOGamblerMem1 was trained, with
probabilities $(\text{Pr}(\text{C}\;|\;\text{CC}),
                \text{Pr}(\text{C}\;|\;\text{CD}),
                \text{Pr}(\text{C}\;|\;\text{DC}),
                \text{Pr}(\text{C}\;|\;\text{DD})) = (1, 0.5217, 0, 0.121)$.
Though it performs well in standard tournaments (see
Table~\ref{tbl:standard_score})
it does not outperform the longer memory strategies, and is bested by a similar
strategy that also uses the first round of play: PSOGambler\_1\_1\_1.

These strategies are trained with a particle swarm algorithm rather than an
evolutionary algorithm (though the former would suffice). Particle swarm
algorithms have been used to trained IPD strategies previously
\cite{franken2005particle}.

\subsection{ANN: Single Hidden Layer Artificial Neural Network}\label{sec:ann}

Strategies based on artificial neural networks use a variety of features
computed from the history of play:

\begin{multicols}{2}
    \begin{itemize}
        \item Opponent's first move is C
        \item Opponent's first move is D
        \item Opponent's second move is C
        \item Opponent's second move is D
        \item Player's previous move is C
        \item Player's previous move is D
        \item Player's second previous move is C
        \item Player's second previous move is D
        \item Opponent's previous move is C
        \item Opponent's previous move is D
        \item Opponent's second previous move is C
        \item Opponent's second previous move is D
        \item Total opponent cooperations
        \item Total opponent defections
        \item Total player cooperations
        \item Total player defections
        \item Round number
    \end{itemize}
\end{multicols}

\begin{figure}[!hbtp]
    \centering
    \begin{tikzpicture}[scale=.9]
    \tikzstyle{cooperator} = [rectangle, pattern=north west lines, pattern
        color=blue!70, draw=blue, text width=.25cm, outer sep = 2pt]
    \tikzstyle{defector} = [rectangle, pattern=north east lines, pattern
        color=red!70, draw=red, text width=.25cm, outer sep = 2pt]

    \node (A) at (-.5, 0) {Player};
    \node (A1) at ($(A) + (0, -.55)$) [cooperator] {C};
    \node (A2) at ($(A1) + (0, -.55)$) [cooperator] {C};
    \node (A3) at ($(A2) + (0, -.55)$) [defector] {D};
    \node (A4) at ($(A3) + (0, -.55)$) [defector] {D};
    \node (A5) at ($(A4) + (0, -.55)$) {$\vdots$};
    \node (A6) at ($(A5) + (0, -.75)$) [defector] {D};
    \node (A7) at ($(A6) + (0, -.55)$) [cooperator] {C};
    \node (A8) at ($(A7) + (0, -.55)$) [cooperator] {C};
    \node (A9) at ($(A8) + (0, -.55)$) [cooperator] {C};
    \node (A10) at ($(A9) + (0, -.55)$) [cooperator] {C};

    \node (B) at ($(A) + (2, 0)$) {Opponent};
    \node (B1) at ($(B) + (0, -.55)$) [cooperator] {C};
    \node (B2) at ($(B1) + (0, -.55)$) [defector] {D};
    \node (B3) at ($(B2) + (0, -.55)$) [cooperator] {C};
    \node (B4) at ($(B3) + (0, -.55)$) [cooperator] {C};
    \node (B5) at ($(B4) + (0, -.55)$) {$\vdots$};
    \node (B6) at ($(B5) + (0, -.75)$) [defector] {D};
    \node (B7) at ($(B6) + (0, -.55)$) [defector] {D};
    \node (B8) at ($(B7) + (0, -.55)$) [defector] {D};
    \node (B9) at ($(B8) + (0, -.55)$) [cooperator] {C};
    \node (B10) at ($(B9) + (0, -.55)$) [cooperator] {C};

    \draw[very thick, black, dashed] ($(A1) + (-.3, .3)$)  rectangle ($(B10) + (.3, -.3)$);

    \node (input_vector) at ($(B5) + (2, 0)$) [right] {(6, 5, 9, \dots, 1)};

    \draw [very thick, black, dashed, ->] ($(B5) + (.3, 0)$) -- (input_vector);


    \node (in1) at ($(input_vector) + (2.5, 1.5)$) [right] {};
    \fill (in1) circle[radius=3pt];
    \node (in2) at ($(input_vector) + (2.5, .5)$) [right] {};
    \fill (in2) circle[radius=3pt];
    \node (in3) at ($(input_vector) + (2.5, -.5)$) [right] {};
    \fill (in3) circle[radius=3pt];

    \node (in4) at ($(input_vector) + (2.625, -1.15)$) {$\vdots$};

    \node (in5) at ($(input_vector) + (2.5, -2)$) [right] {};
    \fill (in5) circle[radius=3pt];


    \draw [->] ($(input_vector) + (-.9, .25)$) to [in=180, out=90] (in1);
    \draw [->] ($(input_vector) + (-.6, .25)$) to [in=180, out=90] (in2);
    \draw [->] ($(input_vector) + (-.2, -.25)$) to [in=180, out=-90] (in3);
    \draw [->] ($(input_vector) + (1, -.25)$) to [in=180, out=-90] (in5);

    \node (mid1) at ($(in1)!0.5!(in2) + (1.5, 0)$) {};
    \fill (mid1) circle[radius=3pt];
    \node (mid2) at ($(in2)!0.5!(in3) + (1.5, 0)$) {};
    \fill (mid2) circle[radius=3pt];
    \node (mid3) at ($(in3)!0.5!(in4) + (1.5, 0)$) {};
    \fill (mid3) circle[radius=3pt];

    \draw [thick] (in1) -- (mid1);
    \draw [thick] (in1) -- (mid2);
    \draw [thick] (in1) -- (mid3);

    \draw [thick] (in2) -- (mid1);
    \draw [thick] (in2) -- (mid2);
    \draw [thick] (in2) -- (mid3);

    \draw [thick] (in3) -- (mid1);
    \draw [thick] (in3) -- (mid2);
    \draw [thick] (in3) -- (mid3);

    \draw [thick] (in5) -- (mid1);
    \draw [thick] (in5) -- (mid2);
    \draw [thick] (in5) -- (mid3);

    \node (output) at ($(mid2) + (2, 0)$) [defector] {D};
    \node (or) at ($(output) + (.6, 0)$) {or};
    \node at ($(or) + (.6, 0)$) [cooperator] {C};

    \draw [->, very thick] (mid1) -- (output);
    \draw [->, very thick] (mid2) -- (output);
    \draw [->, very thick] (mid3) -- (output);
\end{tikzpicture}
    \caption{Diagrammatic representation of the ANN Archetype.}
    \label{fig:ann}
\end{figure}
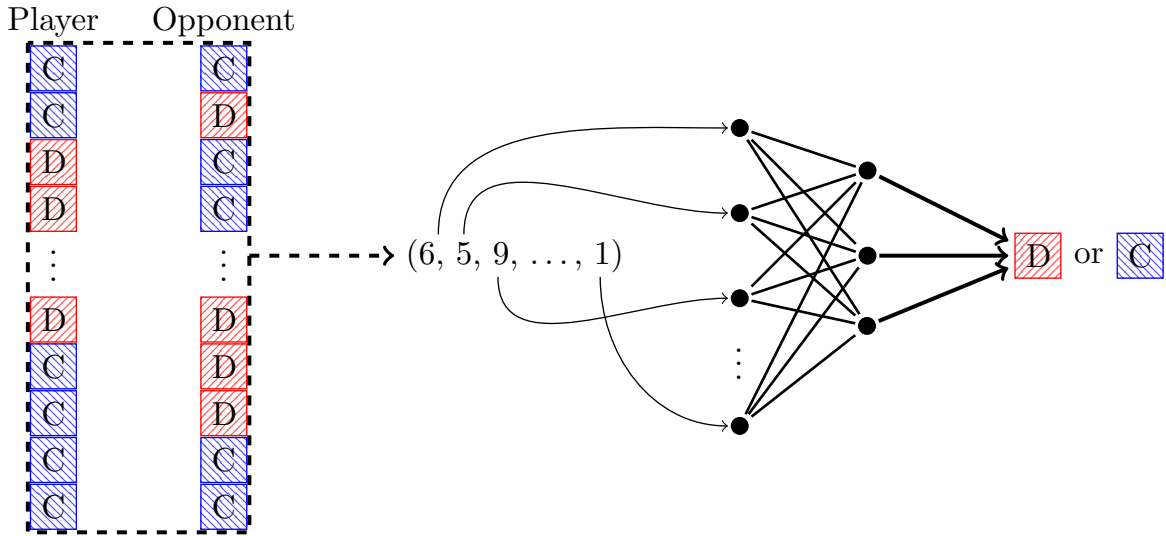

These are then input into a feed forward neural network with one layer and
user-supplied width.  This is illustrated diagrammatically in
Figure~\ref{fig:ann}.

Training of this strategy corresponds to finding parameters of the neural
network. An inner layer with just five nodes performs quite well in both deterministic and
noisy tournaments. The output of the ANN used in this work is deterministic;
a stochastic variant that outputs probabilities rather than exact moves could
be easily created.

\subsection{Finite State Machines}\label{sec:fsm}

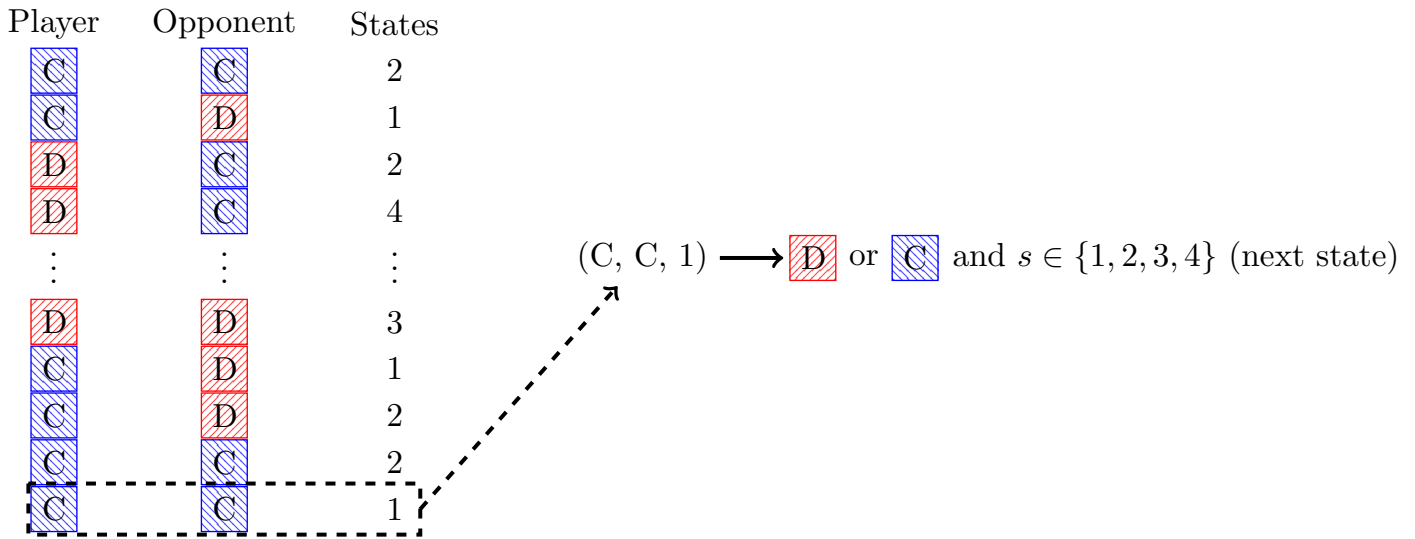
\begin{figure}[!hbtp]
    \centering
    \begin{tikzpicture}[scale=.9]
    \tikzstyle{cooperator} = [rectangle, pattern=north west lines, pattern
        color=blue!70, draw=blue, text width=.25cm, outer sep = 2pt]
    \tikzstyle{defector} = [rectangle, pattern=north east lines, pattern
        color=red!70, draw=red, text width=.25cm, outer sep = 2pt]

    \node (A) at (-.5, 0) {Player};
    \node (A1) at ($(A) + (0, -.55)$) [cooperator] {C};
    \node (A2) at ($(A1) + (0, -.55)$) [cooperator] {C};
    \node (A3) at ($(A2) + (0, -.55)$) [defector] {D};
    \node (A4) at ($(A3) + (0, -.55)$) [defector] {D};
    \node (A5) at ($(A4) + (0, -.55)$) {$\vdots$};
    \node (A6) at ($(A5) + (0, -.75)$) [defector] {D};
    \node (A7) at ($(A6) + (0, -.55)$) [cooperator] {C};
    \node (A8) at ($(A7) + (0, -.55)$) [cooperator] {C};
    \node (A9) at ($(A8) + (0, -.55)$) [cooperator] {C};
    \node (A10) at ($(A9) + (0, -.55)$) [cooperator] {C};

    \node (B) at ($(A) + (2, 0)$) {Opponent};
    \node (B1) at ($(B) + (0, -.55)$) [cooperator] {C};
    \node (B2) at ($(B1) + (0, -.55)$) [defector] {D};
    \node (B3) at ($(B2) + (0, -.55)$) [cooperator] {C};
    \node (B4) at ($(B3) + (0, -.55)$) [cooperator] {C};
    \node (B5) at ($(B4) + (0, -.55)$) {$\vdots$};
    \node (B6) at ($(B5) + (0, -.75)$) [defector] {D};
    \node (B7) at ($(B6) + (0, -.55)$) [defector] {D};
    \node (B8) at ($(B7) + (0, -.55)$) [defector] {D};
    \node (B9) at ($(B8) + (0, -.55)$) [cooperator] {C};
    \node (B10) at ($(B9) + (0, -.55)$) [cooperator] {C};

    \node (S) at ($(B) + (2, 0)$) {States};
    \node (S1) at ($(S) + (0, -.55)$)  {2};
    \node (S2) at ($(S1) + (0, -.55)$)  {1};
    \node (S3) at ($(S2) + (0, -.55)$)  {2};
    \node (S4) at ($(S3) + (0, -.55)$)  {4};
    \node (S5) at ($(S4) + (0, -.55)$) {$\vdots$};
    \node (S6) at ($(S5) + (0, -.75)$)  {3};
    \node (S7) at ($(S6) + (0, -.55)$)  {1};
    \node (S8) at ($(S7) + (0, -.55)$)  {2};
    \node (S9) at ($(S8) + (0, -.55)$)  {2};
    \node (S10) at ($(S9) + (0, -.55)$)  {1};

    \draw[very thick, black, dashed] ($(A10) + (-.3, .3)$)  rectangle ($(S10) + (.3, -.3)$);

    \node (inputs) at ($(S5) + (2, 0)$) [right] {(C, C, 1)};

    \draw [very thick, black, dashed, ->] ($(S10) + (.3, 0)$) -- (inputs);

    \node (output) at ($(inputs) + (2, 0)$) [defector] {D};
    \node (or) at ($(output) + (.6, 0)$) {or};
    \node (c) at ($(or) + (.6, 0)$) [cooperator] {C};
    \node at ($(c) + (0.3, 0)$) [right] {and \(s\in\{1, 2, 3, 4\}\) (next state)};


    \draw [->, very thick] (inputs) -- (output);
\end{tikzpicture}
    \caption{Diagrammatic representation of the Finite State Machine Archetype.}
    \label{fig:fsm}
\end{figure}

Strategies based on finite state machines are deterministic and computationally efficient.
In each round of play the strategy selects an action based on the current state
and the opponent's last action, transitioning to a new state for the next round.
This is illustrated diagrammatically in Figure~\ref{fig:fsm}.

\begin{figure}[!hbtp]
    \centering
    \begin{subfigure}[t]{.5\textwidth}
        \includegraphics[height=.3\textheight]{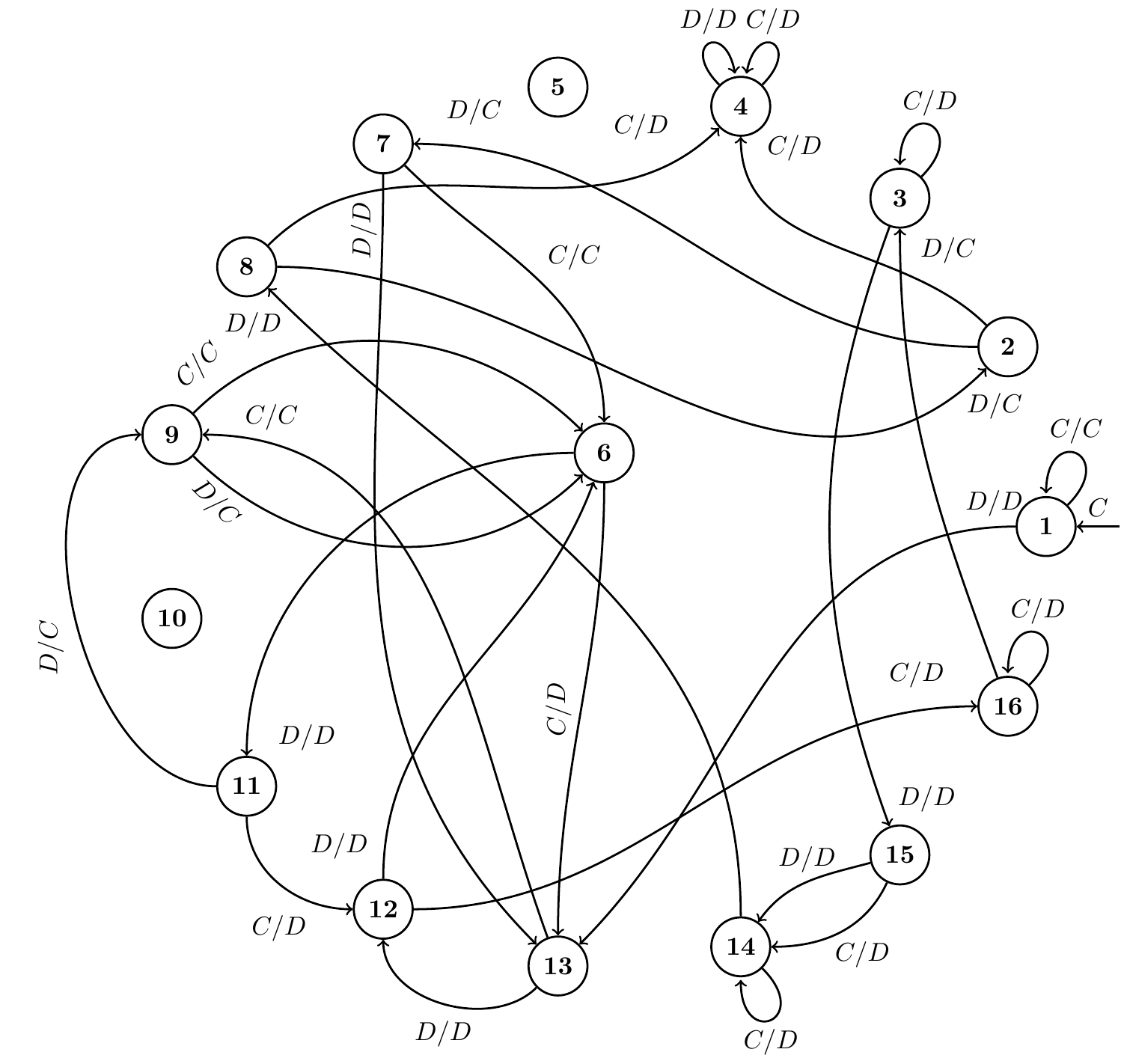}
        \caption{Evolved\_FSM\_16: trained to maximise score in a standard
        tournament}
        \label{fig:fsm16}
    \end{subfigure}%
    ~
    \begin{subfigure}[t]{.5\textwidth}
        \centering
        \includegraphics[height=.35\textheight]{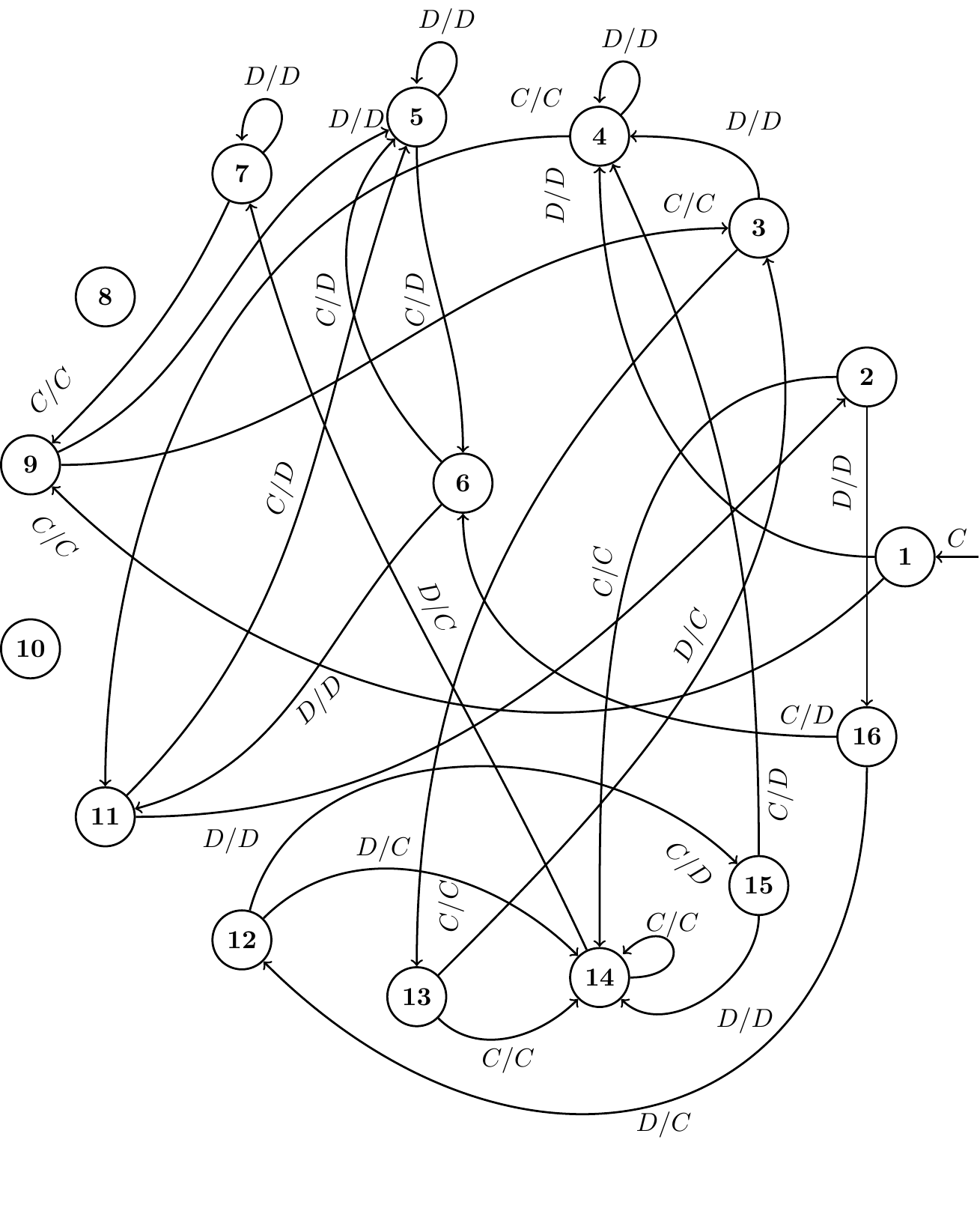}
        \caption{Evolved\_FSM\_16\_Noise\_05: trained to maximise score in a
        noisy tournament}
        \label{fig:fsm16noise}
    \end{subfigure}%
    \caption{Trained sixteen state Finite State Machine players.}
    \label{fig:fsm_images}
\end{figure}

Training this strategy corresponds to finding mappings of states and histories
to an action and a state. Figure~\ref{fig:fsm_images} shows two of the trained
finite state machines. The layout of state nodes is kept the same between
Figure~\ref{fig:fsm16} and~\ref{fig:fsm16noise} to highlight the effect of
different training environments. Note also that two of the 16 states are not
used, this is also an outcome of the training process.

\subsection{Hidden Markov Models}\label{sec:hmm}

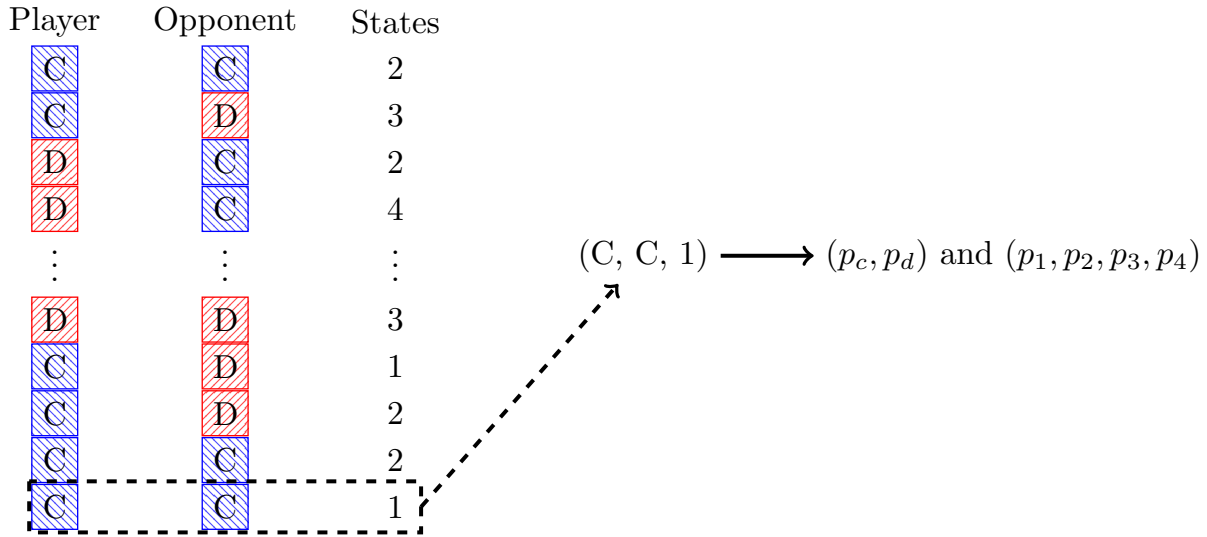
\begin{figure}[!hbtp]
    \centering
    \begin{tikzpicture}[scale=.9]
    \tikzstyle{cooperator} = [rectangle, pattern=north west lines, pattern
        color=blue!70, draw=blue, text width=.25cm]
    \tikzstyle{defector} = [rectangle, pattern=north east lines, pattern
        color=red!70, draw=red, text width=.25cm]

    \node (A) at (-.5, 0) {Player};
    \node (A1) at ($(A) + (0, -.55)$) [cooperator] {C};
    \node (A2) at ($(A1) + (0, -.55)$) [cooperator] {C};
    \node (A3) at ($(A2) + (0, -.55)$) [defector] {D};
    \node (A4) at ($(A3) + (0, -.55)$) [defector] {D};
    \node (A5) at ($(A4) + (0, -.55)$) {$\vdots$};
    \node (A6) at ($(A5) + (0, -.75)$) [defector] {D};
    \node (A7) at ($(A6) + (0, -.55)$) [cooperator] {C};
    \node (A8) at ($(A7) + (0, -.55)$) [cooperator] {C};
    \node (A9) at ($(A8) + (0, -.55)$) [cooperator] {C};
    \node (A10) at ($(A9) + (0, -.55)$) [cooperator] {C};

    \node (B) at ($(A) + (2, 0)$) {Opponent};
    \node (B1) at ($(B) + (0, -.55)$) [cooperator] {C};
    \node (B2) at ($(B1) + (0, -.55)$) [defector] {D};
    \node (B3) at ($(B2) + (0, -.55)$) [cooperator] {C};
    \node (B4) at ($(B3) + (0, -.55)$) [cooperator] {C};
    \node (B5) at ($(B4) + (0, -.55)$) {$\vdots$};
    \node (B6) at ($(B5) + (0, -.75)$) [defector] {D};
    \node (B7) at ($(B6) + (0, -.55)$) [defector] {D};
    \node (B8) at ($(B7) + (0, -.55)$) [defector] {D};
    \node (B9) at ($(B8) + (0, -.55)$) [cooperator] {C};
    \node (B10) at ($(B9) + (0, -.55)$) [cooperator] {C};

    \node (S) at ($(B) + (2, 0)$) {States};
    \node (S1) at ($(S) + (0, -.55)$)  {2};
    \node (S2) at ($(S1) + (0, -.55)$)  {3};
    \node (S3) at ($(S2) + (0, -.55)$)  {2};
    \node (S4) at ($(S3) + (0, -.55)$)  {4};
    \node (S5) at ($(S4) + (0, -.55)$) {$\vdots$};
    \node (S6) at ($(S5) + (0, -.75)$)  {3};
    \node (S7) at ($(S6) + (0, -.55)$)  {1};
    \node (S8) at ($(S7) + (0, -.55)$)  {2};
    \node (S9) at ($(S8) + (0, -.55)$)  {2};
    \node (S10) at ($(S9) + (0, -.55)$)  {1};

    \draw[very thick, black, dashed] ($(A10) + (-.3, .3)$)  rectangle ($(S10) + (.3, -.3)$);

    \node (inputs) at ($(S5) + (2, 0)$) [right] {(C, C, 1)};

    \draw [very thick, black, dashed, ->] ($(S10) + (.3, 0)$) -- (inputs);

    \node (output) at ($(inputs) + (2, 0)$) [right] {\((p_c, p_d)\) and \((p_1, p_2, p_3, p_4)\)};

    \draw [->, very thick] (inputs) -- (output);
\end{tikzpicture}
    \caption{Diagrammatic representation of the Hidden Markov Model Archetype.}
    \label{fig:hmm}
\end{figure}

A variant of finite state machine strategies are called hidden Markov models
(HMMs). Like the strategies based on finite state machines, these strategies
also encode an internal state. However, they use probabilistic transitions based on the
prior round of play to other states and cooperate or defect with various
probabilities at each state. This is
shown diagrammatically in Figure~\ref{fig:hmm}. Training this strategy
corresponds to finding mappings of states and histories to probabilities of
cooperating as well as probabilities of the next internal state.

\subsection{Meta Strategies}

There are several strategies based on ensemble methods that
are common in machine learning called Meta strategies. These strategies are
composed of a team of other strategies. In each round, each member of the team
is polled for its desired next
move. The ensemble then selects the next move based on a rule, such as the
consensus vote in the case of MetaMajority or the best individual performance
in the case of MetaWinner. These strategies were among the best in the library
before the inclusion of those trained by reinforcement learning. The library
contains strategies containing teams of all the deterministic players, all the
memory-one players, and some others.

Because these strategies inherit many of the properties of the strategies
on which they are based, including using knowledge of the match length to defect
on the last round(s) of play, not all of these
strategies were included in results of this
paper. These strategies do not typically outperform the trained strategies
described above.

\section{Results}\label{sec:results}

This section presents the results of a large IPD tournament with
strategies from the Axelrod library, including some additional parametrized
strategies (e.g.\ various parameter choices for Generous Tit For Tat
\cite{Gaudesi2016}). These are
listed in Appendix~\ref{app:list_of_players}.

All strategies in the tournament follow a simple set of
rules in accordance with earlier tournaments:

\begin{itemize}
  \item Players are unaware of the number of turns in a match.
  \item Players carry no acquired state between matches.
  \item Players cannot observe the outcome of other matches.
  \item Players cannot identify their opponent by any label or identifier.
  \item Players cannot manipulate or inspect their opponents in any way.
\end{itemize}

Any strategy that does not follow these rules, such as a strategy that defects
on the last round of play, was omitted from the tournament presented here (but
not necessarily from the training pool).

\begin{figure}[!hbtp]
    \centering
    \begin{subfigure}[t]{.3\textwidth}
        \centering
        \includegraphics[width=\textwidth]{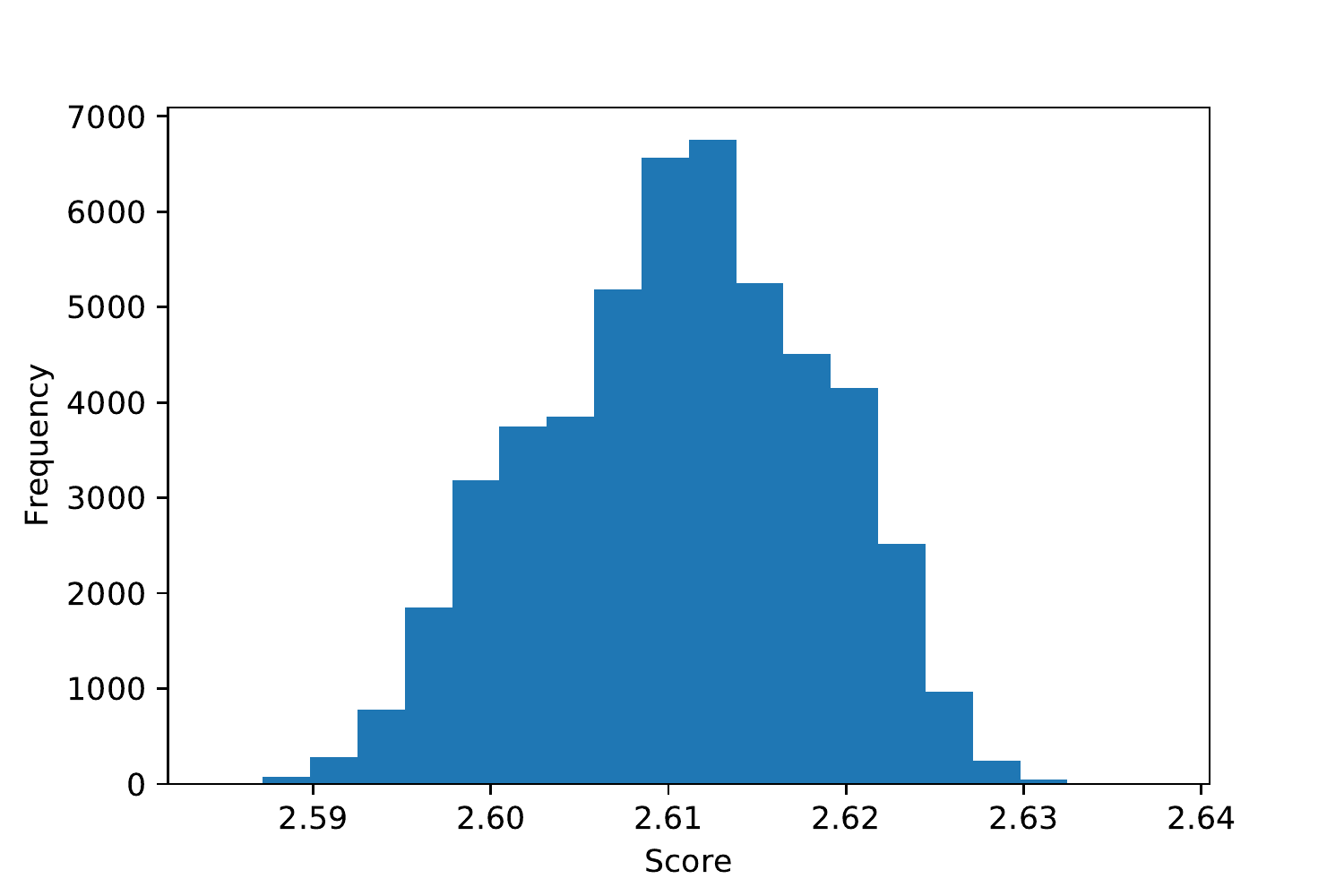}
        \caption{Scores}
        \label{fig:tit_for_tat_scores}
    \end{subfigure}%
    ~
    \begin{subfigure}[t]{.3\textwidth}
        \centering
        \includegraphics[width=\textwidth]{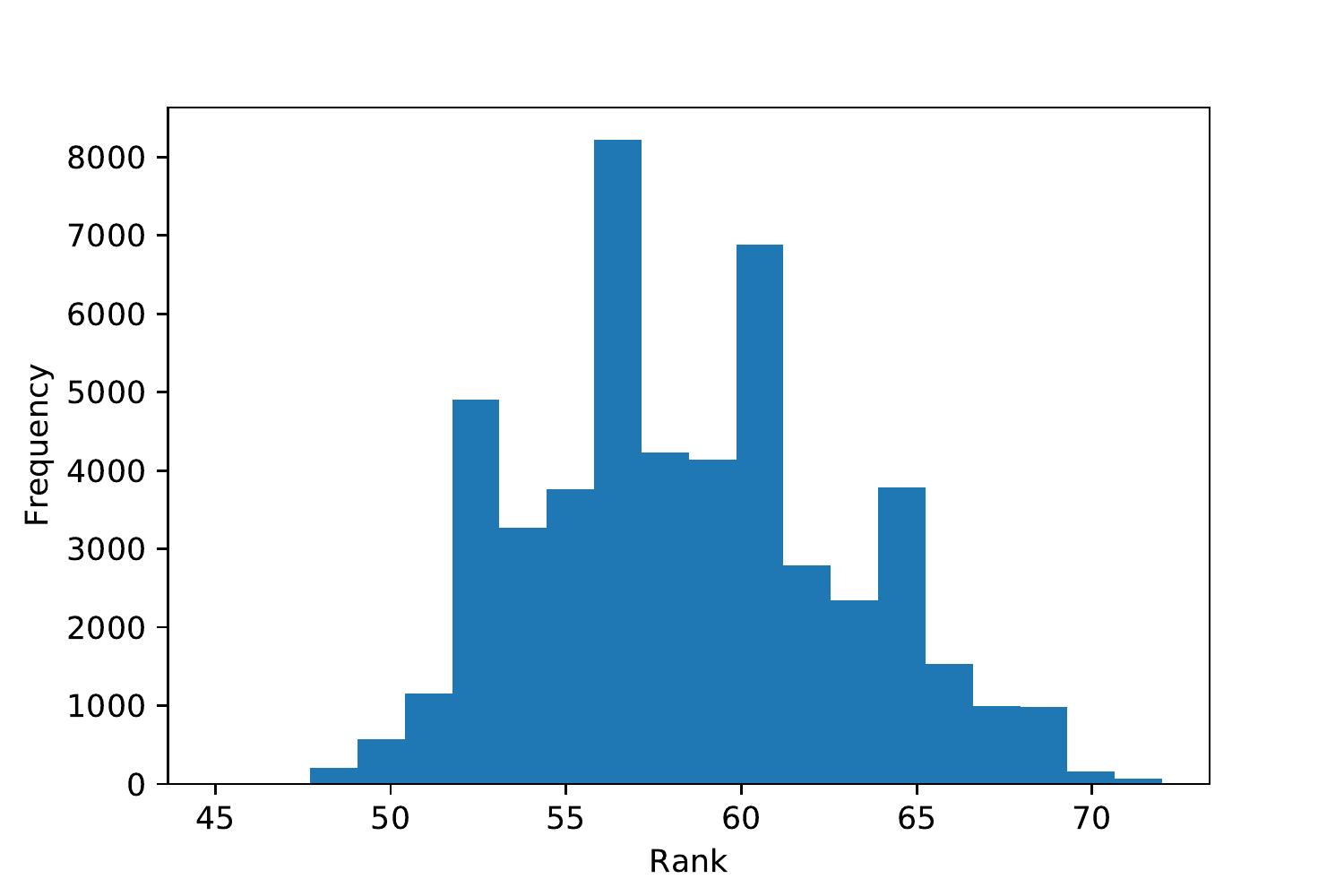}
        \caption{Ranks}
        \label{fig:tit_for_tat_ranks}
    \end{subfigure}%
    ~
    \begin{subfigure}[t]{.3\textwidth}
        \centering
        \includegraphics[width=\textwidth]{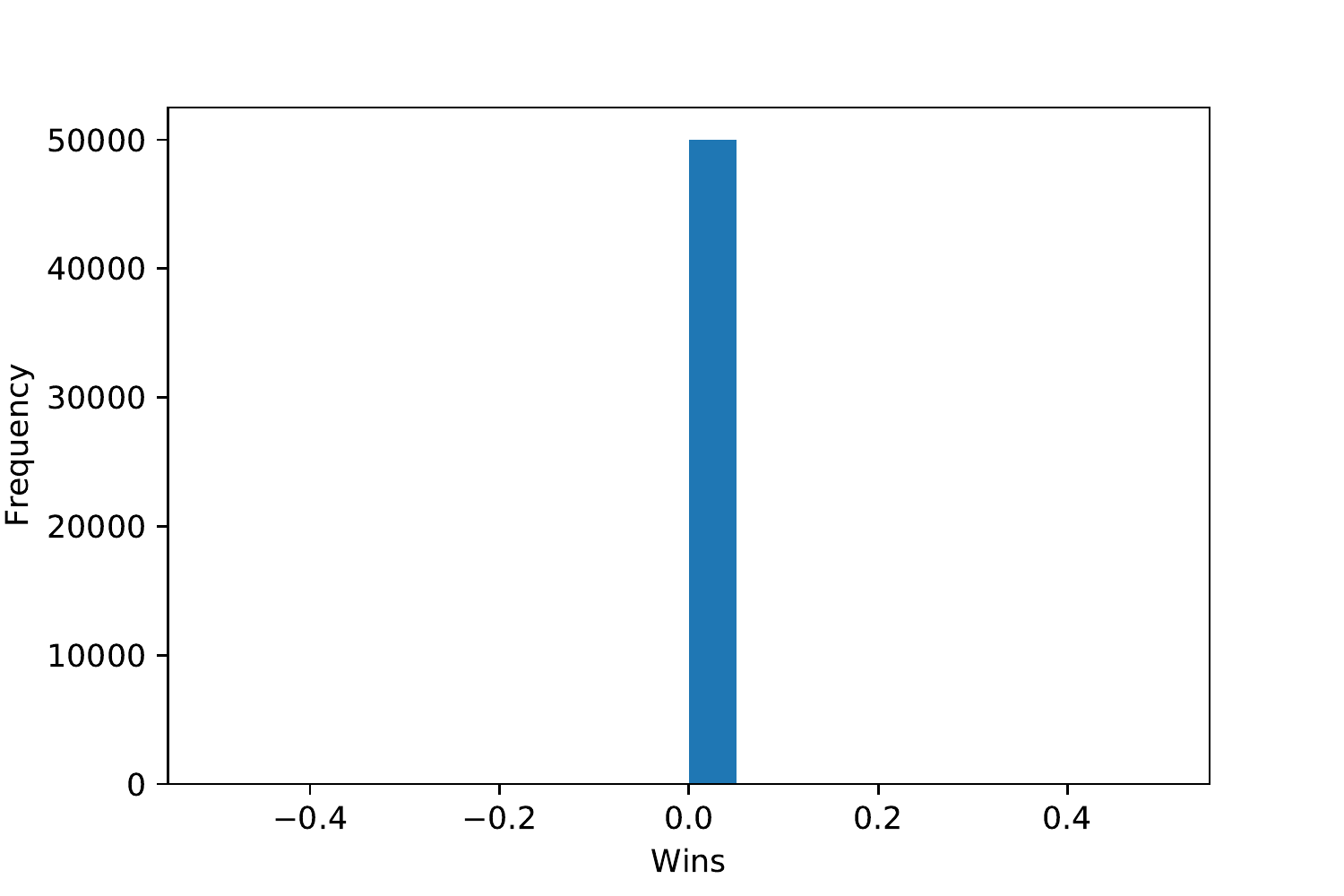}
        \caption{Wins}
        \label{fig:tit_for_tat_wins}
    \end{subfigure}%

    \caption{Results for Tit For Tat over
        \protect\input{./assets/standard_number_of_repetitions.tex}tournaments.}
\end{figure}

A total of 176are included, of which
53are stochastic.  In
Section~\ref{sec:standard} is concerned with the standard tournament with 200
turns whereas in Section~\ref{sec:noise} a tournament with 5\% noise is
discussed. Due to the inherent stochasticity of these IPD tournaments, these
tournament were repeated 50000
times. This allows for a detailed and confident analysis of the performance of
strategies. To illustrate the results considered,
Figure~\ref{fig:tit_for_tat_scores} shows the distribution of the mean score per
turn of Tit For Tat over all the repetitions. Similarly,
Figure~\ref{fig:tit_for_tat_ranks} shows the ranks of of Tit For Tat for each
repetition (we note that it never wins a tournament). Finally
Figure~\ref{fig:tit_for_tat_wins} shows the number of opponents beaten in any given
tournament: Tit For Tat does not win any match (this is due to the fact that it
will either draw with mutual cooperation or defect second).

The utilities used are \((R, P, T, S)=(3, 1, 5, 0)\) thus the specific
Prisoner's Dilemma being played is:

\begin{equation}\label{equ:pd}
    \begin{pmatrix}
        (3, 3) & (0, 5)\\
        (5, 0) & (1, 1)
    \end{pmatrix}
\end{equation}

All data generated for this work is archived and available at~\cite{data}.

\subsection{Standard Tournament}\label{sec:standard}

The top 11 performing strategies by median payoff are all strategies trained to maximize
total payoff against a subset of the strategies (Table~\ref{tbl:standard_score}).
The next strategy is Desired Belief Strategy (DBS) \cite{Au2006},
which actively analyzes the opponent and responds
accordingly. The next two strategies are Winner12, based on a lookup table,
Fool Me Once \cite{axelrodproject}, a grudging strategy that defects indefinitely on
the second defection, and Omega Tit For Tat \cite{kendall2007iterated}.

\begin{table}[!hbtp]
        \centering
        \begin{tabular}{lrrrrrrrrr}
\toprule
{} &   mean &    std &    min &     5\% &    25\% &    50\% &    75\% &    95\% &    max \\
\midrule
EvolvedLookerUp2\_2\_2$^{*}$    &  2.955 &  0.010 &  2.915 &  2.937 &  2.948 &  2.956 &  2.963 &  2.971 &  2.989 \\
Evolved HMM 5$^{*}$           &  2.954 &  0.014 &  2.903 &  2.931 &  2.945 &  2.954 &  2.964 &  2.977 &  3.007 \\
Evolved FSM 16$^{*}$          &  2.952 &  0.013 &  2.900 &  2.930 &  2.943 &  2.953 &  2.962 &  2.973 &  2.993 \\
PSO Gambler 2\_2\_2$^{*}$       &  2.938 &  0.013 &  2.884 &  2.914 &  2.930 &  2.940 &  2.948 &  2.957 &  2.972 \\
Evolved FSM 16 Noise 05$^{*}$ &  2.919 &  0.013 &  2.874 &  2.898 &  2.910 &  2.919 &  2.928 &  2.939 &  2.965 \\
PSO Gambler 1\_1\_1$^{*}$       &  2.912 &  0.023 &  2.805 &  2.874 &  2.896 &  2.912 &  2.928 &  2.950 &  3.012 \\
Evolved ANN 5$^{*}$           &  2.912 &  0.010 &  2.871 &  2.894 &  2.905 &  2.912 &  2.919 &  2.928 &  2.945 \\
Evolved FSM 4$^{*}$           &  2.910 &  0.012 &  2.867 &  2.889 &  2.901 &  2.910 &  2.918 &  2.929 &  2.943 \\
Evolved ANN$^{*}$             &  2.907 &  0.010 &  2.865 &  2.890 &  2.900 &  2.908 &  2.914 &  2.923 &  2.942 \\
PSO Gambler Mem1$^{*}$        &  2.901 &  0.025 &  2.783 &  2.858 &  2.884 &  2.901 &  2.919 &  2.942 &  2.994 \\
Evolved ANN 5 Noise 05$^{*}$  &  2.864 &  0.008 &  2.830 &  2.850 &  2.858 &  2.865 &  2.870 &  2.877 &  2.891 \\
DBS                           &  2.857 &  0.009 &  2.823 &  2.842 &  2.851 &  2.857 &  2.863 &  2.872 &  2.899 \\
Winner12                      &  2.849 &  0.008 &  2.820 &  2.836 &  2.844 &  2.850 &  2.855 &  2.862 &  2.874 \\
Fool Me Once                  &  2.844 &  0.008 &  2.818 &  2.830 &  2.838 &  2.844 &  2.850 &  2.857 &  2.882 \\
Omega TFT: 3, 8               &  2.841 &  0.011 &  2.800 &  2.822 &  2.833 &  2.841 &  2.849 &  2.859 &  2.882 \\
\bottomrule
\end{tabular}

        \caption{Standard Tournament: Mean score per turn of top 15 strategies
            (ranked by median over
        \protect 50000tournaments).
        The leaderboard is dominated by the trained strategies (indicated by a
        $^{*}$).}
        \label{tbl:standard_score}
\end{table}

\begin{landscape}
    \begin{figure}[!hbtp]
        \centering
        \includegraphics[width=\paperwidth]{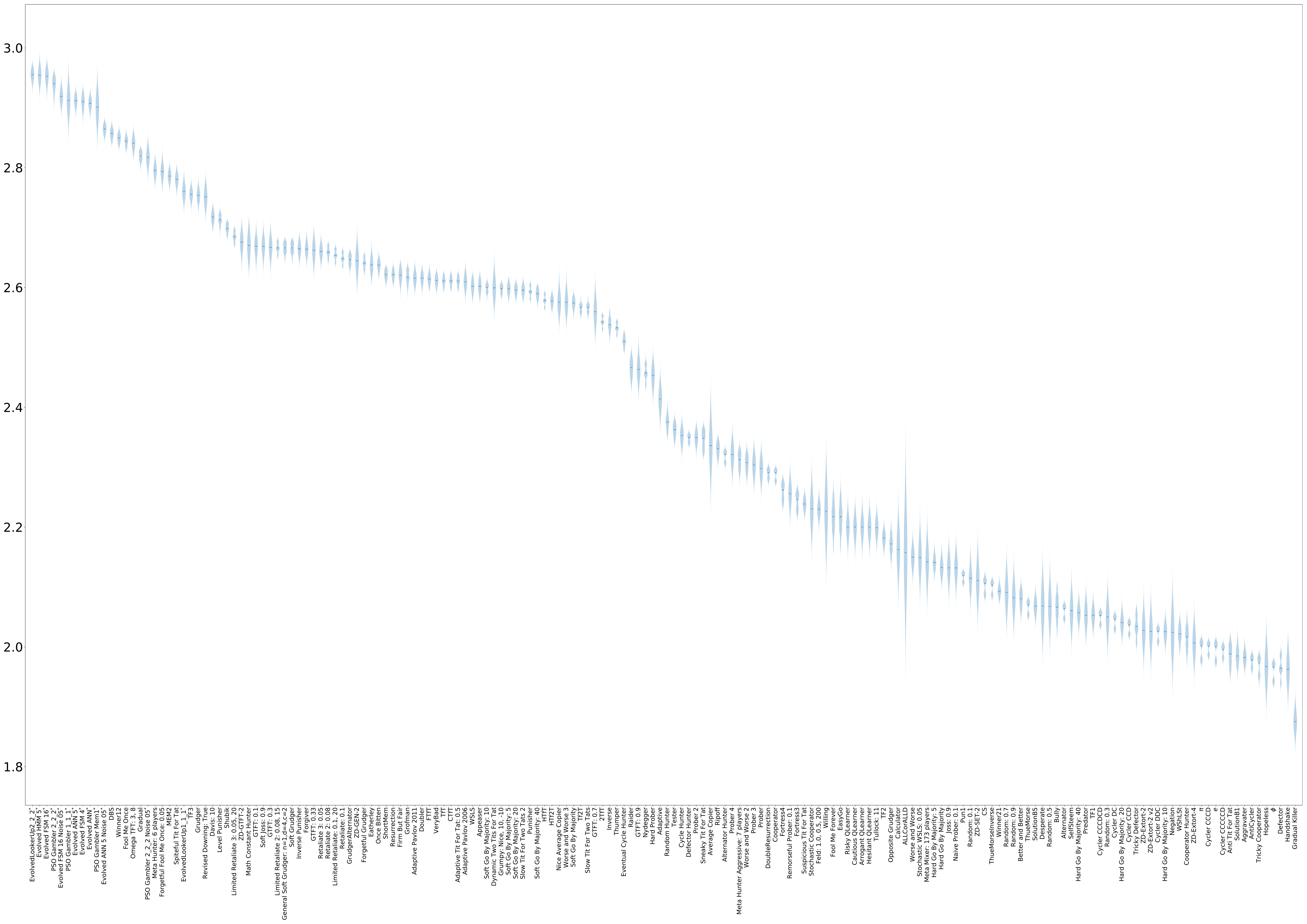}
        \caption{Standard Tournament: Mean score per turn (ranked by median
        over
        \protect\input{./assets/standard_number_of_repetitions.tex}tournaments).}
        \label{fig:standard_boxplot}
    \end{figure}
\end{landscape}

For completeness, violin plots showing the distribution of the scores of each
strategy (again ranked by median score) are shown in
Figure~\ref{fig:standard_boxplot}.

\begin{figure}[!hbtp]
    \centering
    \includegraphics[width=\textwidth]{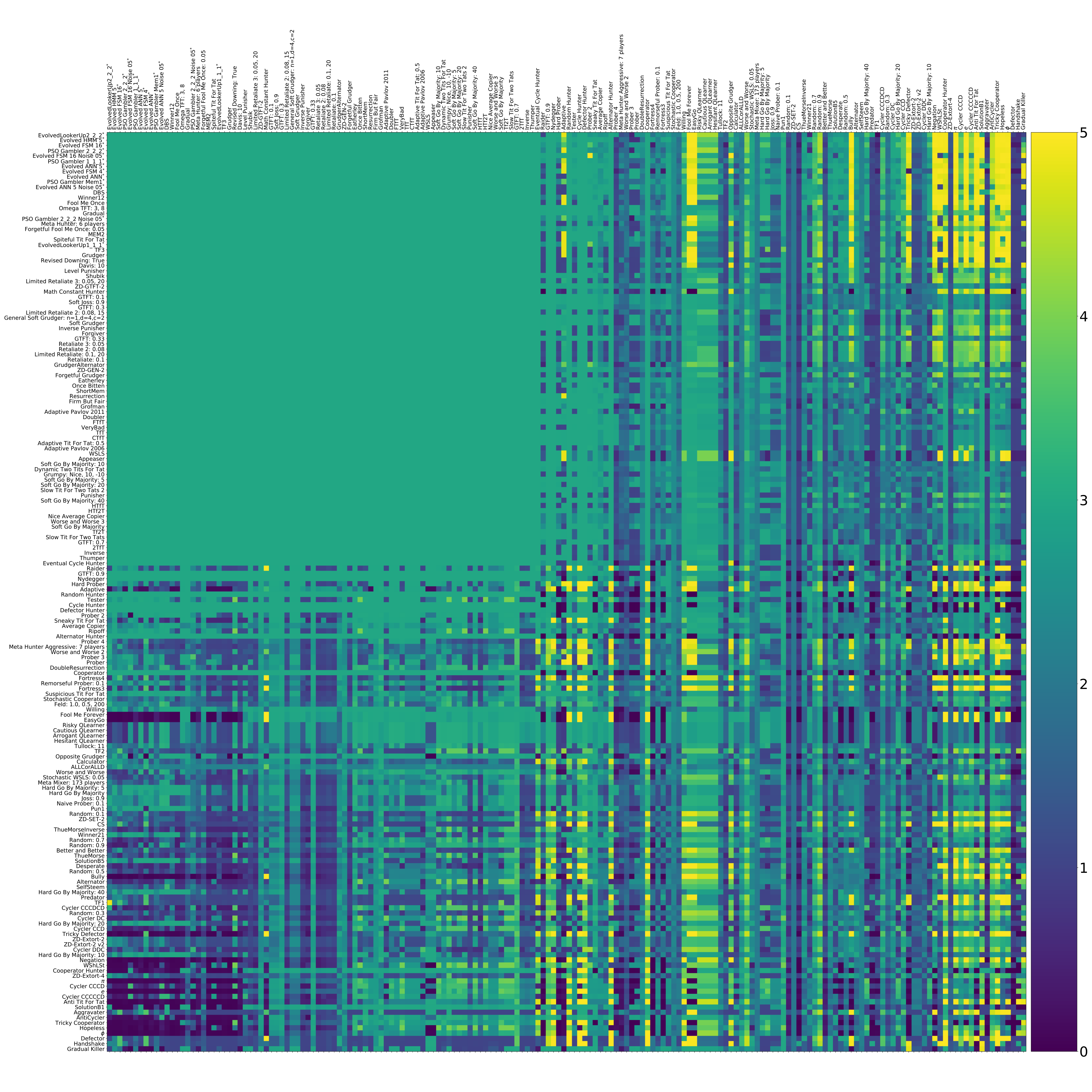}
    \caption{Standard Tournament: Mean score per turn of row players against
    column players (ranked by median over
        \protect\input{./assets/standard_number_of_repetitions.tex}tournaments).}
    \label{fig:standard_heatmap}
\end{figure}

Pairwise payoff results are given as a heatmap (Figure~\ref{fig:standard_heatmap})
which shows that many strategies achieve mutual cooperation (obtaining a score
of 3). The top performing
strategies never defect first yet are able to exploit weaker strategies that
attempt to defect.

\begin{landscape}
    \begin{figure}[!hbtp]
        \centering
        \includegraphics[width=\paperwidth]{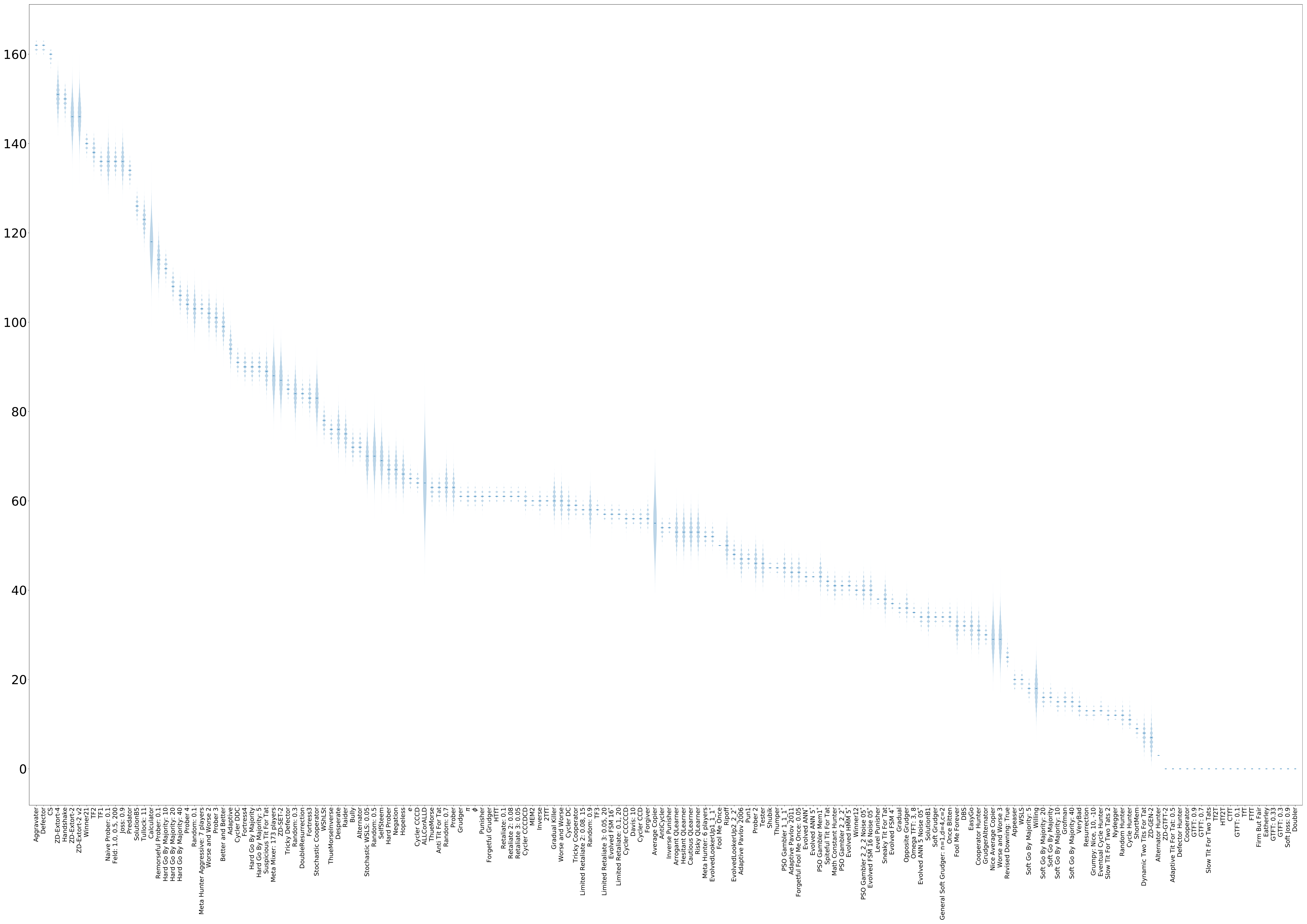}
        \caption{Standard Tournament: number of wins per tournament (ranked by
        median over
        \protect\input{./assets/standard_number_of_repetitions.tex}tournaments).}
        \label{fig:standard_winplot}
    \end{figure}
\end{landscape}

The strategies that win the most matches
(Table~\ref{tbl:standard_wins_top_winners}) are Defector~\cite{Axelrod1984} and Aggravater~\cite{axelrodproject}, followed
by handshaking and zero determinant strategies~\cite{Press2012}.
This includes two handshaking
strategies that were the result of training to maximize Moran process fixation
(TF1 and TF2). No strategies were trained specifically to win matches. None of
the top scoring strategies appear in the top 15 list of strategies ranked by
match wins. This can be seen in Figure~\ref{fig:standard_winplot} where the
distribution of the number of wins of each strategy is shown.

\begin{table}[!hbtp]
    \centering
        \begin{tabular}{lrrrrrrrrr}
\toprule
{} &     mean &    std &  min &     5\% &    25\% &    50\% &    75\% &    95\% &  max \\
\midrule
Aggravater          &  161.595 &  0.862 &  160 &  160.0 &  161.0 &  162.0 &  162.0 &  163.0 &  163 \\
Defector            &  161.605 &  0.864 &  160 &  160.0 &  161.0 &  162.0 &  162.0 &  163.0 &  163 \\
CS                  &  159.646 &  1.005 &  155 &  158.0 &  159.0 &  160.0 &  160.0 &  161.0 &  161 \\
ZD-Extort-4         &  150.598 &  2.662 &  138 &  146.0 &  149.0 &  151.0 &  152.0 &  155.0 &  162 \\
Handshake           &  149.552 &  1.754 &  142 &  147.0 &  148.0 &  150.0 &  151.0 &  152.0 &  154 \\
ZD-Extort-2         &  146.094 &  3.445 &  129 &  140.0 &  144.0 &  146.0 &  148.0 &  152.0 &  160 \\
ZD-Extort-2 v2      &  146.291 &  3.425 &  131 &  141.0 &  144.0 &  146.0 &  149.0 &  152.0 &  160 \\
Winner21            &  139.946 &  1.225 &  136 &  138.0 &  139.0 &  140.0 &  141.0 &  142.0 &  143 \\
TF2                 &  138.240 &  1.700 &  130 &  135.0 &  137.0 &  138.0 &  139.0 &  141.0 &  143 \\
TF1                 &  135.692 &  1.408 &  130 &  133.0 &  135.0 &  136.0 &  137.0 &  138.0 &  140 \\
Naive Prober: 0.1   &  136.016 &  2.504 &  127 &  132.0 &  134.0 &  136.0 &  138.0 &  140.0 &  147 \\
Feld: 1.0, 0.5, 200 &  136.087 &  1.696 &  130 &  133.0 &  135.0 &  136.0 &  137.0 &  139.0 &  144 \\
Joss: 0.9           &  136.015 &  2.503 &  126 &  132.0 &  134.0 &  136.0 &  138.0 &  140.0 &  146 \\
Predator            &  133.718 &  1.385 &  129 &  131.0 &  133.0 &  134.0 &  135.0 &  136.0 &  138 \\
SolutionB5          &  125.843 &  1.509 &  120 &  123.0 &  125.0 &  126.0 &  127.0 &  128.0 &  131 \\
\bottomrule
\end{tabular}

        \caption{Standard Tournament: Number of wins per tournament
        of top 15 strategies (ranked by median wins over
        \protect 50000tournaments).}
        \label{tbl:standard_wins_top_winners}
\end{table}

The number of wins of the top strategies of Table~\ref{tbl:standard_score} are
shown in Table~\ref{tbl:standard_wins}. It is evident that although these
strategies score highly they do not win many matches: the strategy with the most
number of wins is the Evolved FSM 16 strategy that at most won 60
(\(60/175\approx34\%\)) matches in a given tournament.

\begin{table}[!hbtp]
    \centering
        \begin{tabular}{lrrrrrrrrr}
\toprule
{} &    mean &    std &  min &    5\% &   25\% &   50\% &   75\% &   95\% &  max \\
\midrule
EvolvedLookerUp2\_2\_2$^{*}$    &  48.259 &  1.336 &   43 &  46.0 &  47.0 &  48.0 &  49.0 &  50.0 &   53 \\
Evolved HMM 5$^{*}$           &  41.358 &  1.221 &   36 &  39.0 &  41.0 &  41.0 &  42.0 &  43.0 &   45 \\
Evolved FSM 16$^{*}$          &  56.978 &  1.099 &   51 &  55.0 &  56.0 &  57.0 &  58.0 &  59.0 &   60 \\
PSO Gambler 2\_2\_2$^{*}$       &  40.692 &  1.089 &   36 &  39.0 &  40.0 &  41.0 &  41.0 &  42.0 &   45 \\
Evolved FSM 16 Noise 05$^{*}$ &  40.070 &  1.673 &   34 &  37.0 &  39.0 &  40.0 &  41.0 &  43.0 &   47 \\
PSO Gambler 1\_1\_1$^{*}$       &  45.005 &  1.595 &   38 &  42.0 &  44.0 &  45.0 &  46.0 &  48.0 &   51 \\
Evolved ANN 5$^{*}$           &  43.224 &  0.674 &   41 &  42.0 &  43.0 &  43.0 &  44.0 &  44.0 &   47 \\
Evolved FSM 4$^{*}$           &  37.227 &  0.951 &   34 &  36.0 &  37.0 &  37.0 &  38.0 &  39.0 &   41 \\
Evolved ANN$^{*}$             &  43.100 &  1.021 &   40 &  42.0 &  42.0 &  43.0 &  44.0 &  45.0 &   48 \\
PSO Gambler Mem1$^{*}$        &  43.444 &  1.837 &   34 &  40.0 &  42.0 &  43.0 &  45.0 &  46.0 &   51 \\
Evolved ANN 5 Noise 05$^{*}$  &  33.711 &  1.125 &   30 &  32.0 &  33.0 &  34.0 &  34.0 &  35.0 &   38 \\
DBS                           &  32.329 &  1.198 &   28 &  30.0 &  32.0 &  32.0 &  33.0 &  34.0 &   38 \\
Winner12                      &  40.179 &  1.037 &   36 &  39.0 &  39.0 &  40.0 &  41.0 &  42.0 &   44 \\
Fool Me Once                  &  50.121 &  0.422 &   48 &  50.0 &  50.0 &  50.0 &  50.0 &  51.0 &   52 \\
Omega TFT: 3, 8               &  35.157 &  0.859 &   32 &  34.0 &  35.0 &  35.0 &  36.0 &  37.0 &   39 \\
\bottomrule
\end{tabular}

        \caption{Standard Tournament: Number of wins per tournament
        of top 15 strategies (ranked by median score over
        \protect 50000tournaments).}
        \label{tbl:standard_wins}
\end{table}

\begin{landscape}
    \begin{figure}[!hbtp]
        \centering
        \includegraphics[width=\paperwidth]{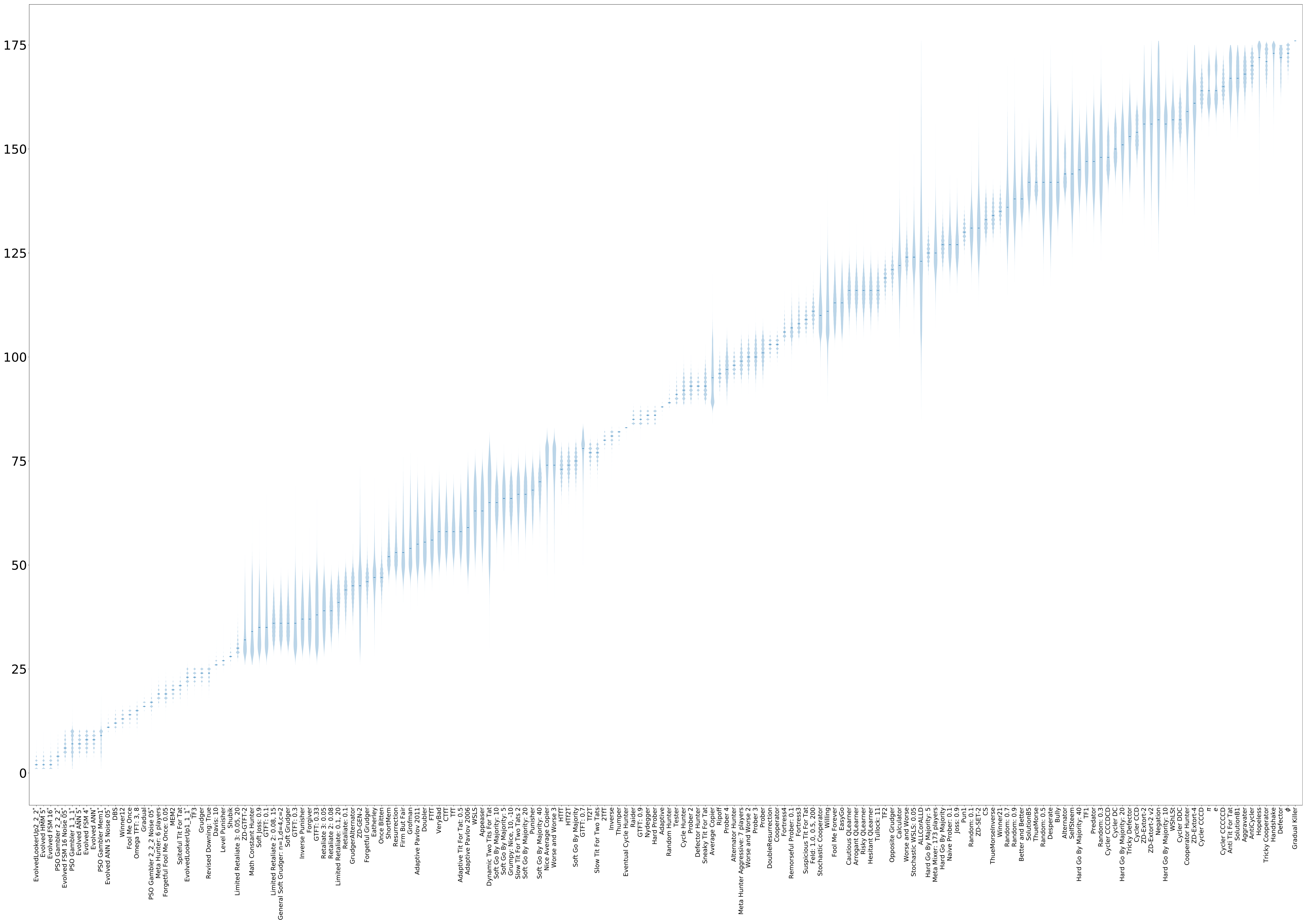}
        \caption{Standard Tournament: rank in each tournament (ranked by
        median over
        \protect\input{./assets/standard_number_of_repetitions.tex}tournaments).}
        \label{fig:standard_ranks_boxplot}
    \end{figure}
\end{landscape}

Finally, Table~\ref{tbl:standard_ranks} and
Figure~\ref{fig:standard_ranks_boxplot} show the ranks (based on median score)
of each strategy over the repeated tournaments. Whilst there is some
stochasticity, the top three strategies almost always rank in the top three. For
example, the worst that the Evolved Lookerup 2 2 2 ranks in any tournament
is 8th.

\begin{table}[!hbtp]
    \centering
        \begin{tabular}{lrrrrrrrrr}
\toprule
{} &    mean &    std &  min &    5\% &   25\% &   50\% &   75\% &   95\% &  max \\
\midrule
EvolvedLookerUp2\_2\_2$^{*}$    &   2.173 &  1.070 &    1 &   1.0 &   1.0 &   2.0 &   3.0 &   4.0 &    8 \\
Evolved HMM 5$^{*}$           &   2.321 &  1.275 &    1 &   1.0 &   1.0 &   2.0 &   3.0 &   5.0 &   10 \\
Evolved FSM 16$^{*}$          &   2.489 &  1.299 &    1 &   1.0 &   1.0 &   2.0 &   3.0 &   5.0 &   10 \\
PSO Gambler 2\_2\_2$^{*}$       &   3.961 &  1.525 &    1 &   2.0 &   3.0 &   4.0 &   5.0 &   7.0 &   10 \\
Evolved FSM 16 Noise 05$^{*}$ &   6.300 &  1.688 &    1 &   4.0 &   5.0 &   6.0 &   7.0 &   9.0 &   11 \\
PSO Gambler 1\_1\_1$^{*}$       &   7.082 &  2.499 &    1 &   3.0 &   5.0 &   7.0 &   9.0 &  10.0 &   17 \\
Evolved ANN 5$^{*}$           &   7.287 &  1.523 &    2 &   5.0 &   6.0 &   7.0 &   8.0 &  10.0 &   11 \\
Evolved FSM 4$^{*}$           &   7.527 &  1.631 &    2 &   5.0 &   6.0 &   8.0 &   9.0 &  10.0 &   12 \\
Evolved ANN$^{*}$             &   7.901 &  1.450 &    2 &   5.0 &   7.0 &   8.0 &   9.0 &  10.0 &   12 \\
PSO Gambler Mem1$^{*}$        &   8.222 &  2.535 &    1 &   4.0 &   6.0 &   9.0 &  10.0 &  12.0 &   20 \\
Evolved ANN 5 Noise 05$^{*}$  &  11.362 &  0.872 &    8 &  10.0 &  11.0 &  11.0 &  12.0 &  13.0 &   16 \\
DBS                           &  12.197 &  1.125 &    9 &  11.0 &  11.0 &  12.0 &  13.0 &  14.0 &   16 \\
Winner12                      &  13.221 &  1.137 &    9 &  11.0 &  12.0 &  13.0 &  14.0 &  15.0 &   17 \\
Fool Me Once                  &  13.960 &  1.083 &    9 &  12.0 &  13.0 &  14.0 &  15.0 &  15.0 &   17 \\
Omega TFT: 3, 8               &  14.275 &  1.301 &    9 &  12.0 &  13.0 &  15.0 &  15.0 &  16.0 &   19 \\
\bottomrule
\end{tabular}

        \caption{Standard Tournament: Rank in each tournament
        of top 15 strategies (ranked by median over
        \protect 50000tournaments).}
        \label{tbl:standard_ranks}
\end{table}

\begin{figure}[!hbtp]
    \centering
    \begin{subfigure}[t]{.3\textwidth}
        \centering
        \includegraphics[width=\textwidth]{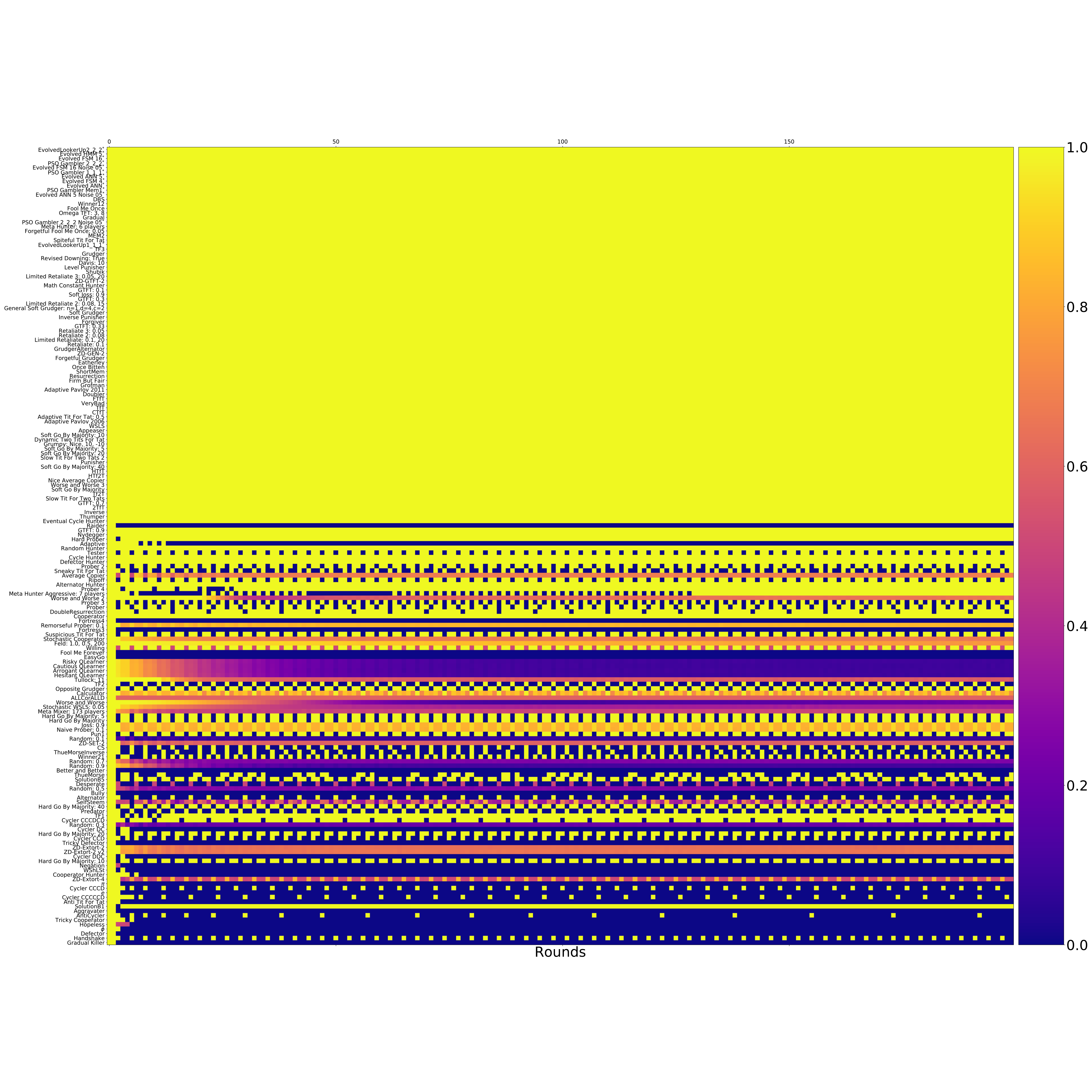}
        \caption{EvolvedLookerUp\_2\_2\_2}
    \end{subfigure}%
    ~
    \begin{subfigure}[t]{.3\textwidth}
        \centering
        \includegraphics[width=\textwidth]{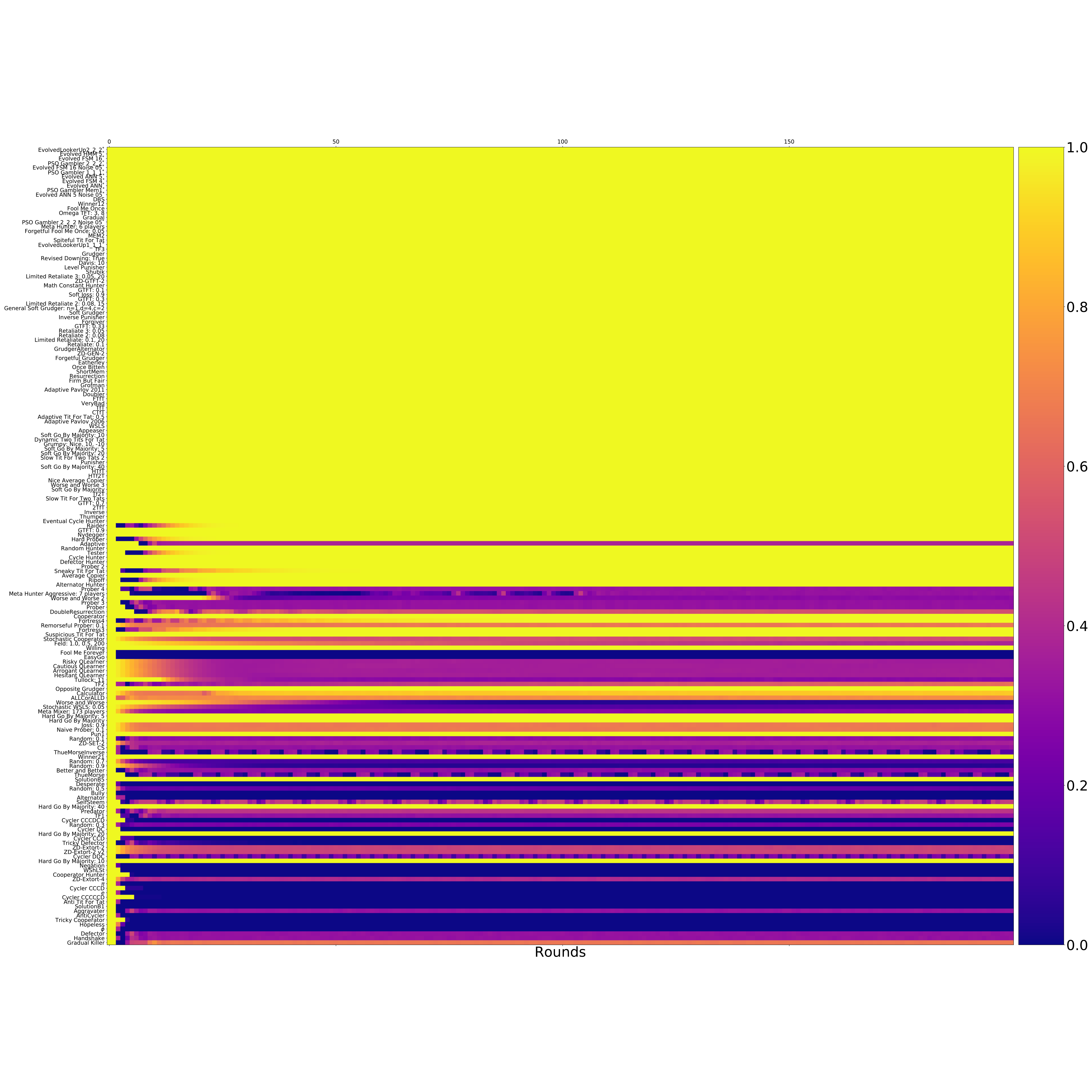}
        \caption{Evolved\_HMM\_5}
    \end{subfigure}%
    ~
    \begin{subfigure}[t]{.3\textwidth}
        \centering
        \includegraphics[width=\textwidth]{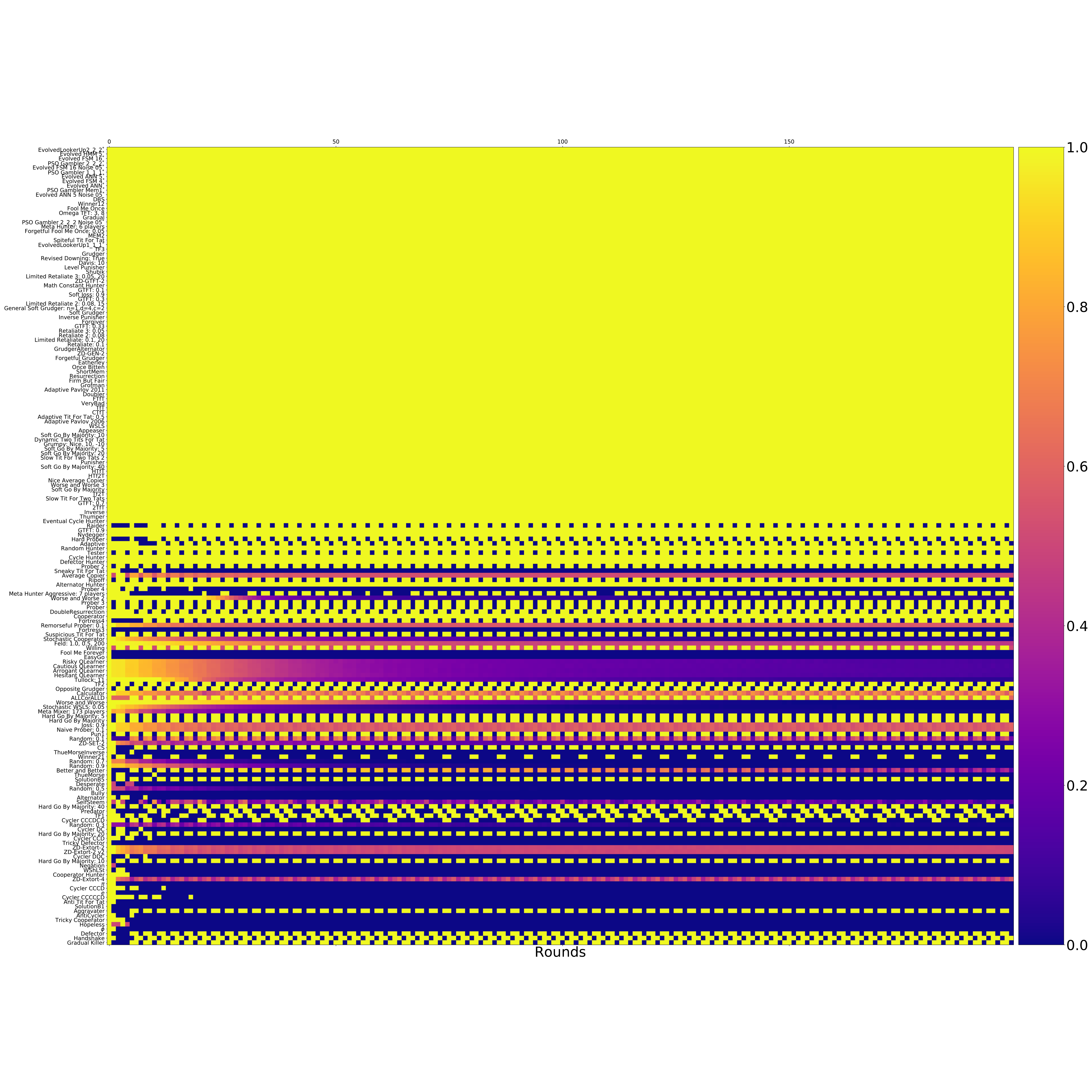}
        \caption{Evolved\_FSM\_16}
    \end{subfigure}%

    \caption{Comparison of cooperation rates for Standard Tournament Top 3
    (over 10000 repetitions).}
    \label{fig:comparison_cooperation_heatmaps_standard}
\end{figure}

Figure~\ref{fig:comparison_cooperation_heatmaps_standard} shows the rate of
cooperation in each round for the top three strategies. The opponents in these
figures are ordered according to performance by median score. It is evident that
the high performing strategies share a common thread against the top strategies:
they do not defect first and achieve mutual cooperation. Against the lower
strategies they also do not defect first (a mean cooperation rate of 1 in the
first round) but do learn to quickly retaliate.

\subsection{Noisy Tournament}\label{sec:noise}

\begin{landscape}
    \begin{figure}[!hbtp]
        \centering
        \includegraphics[width=\paperwidth]{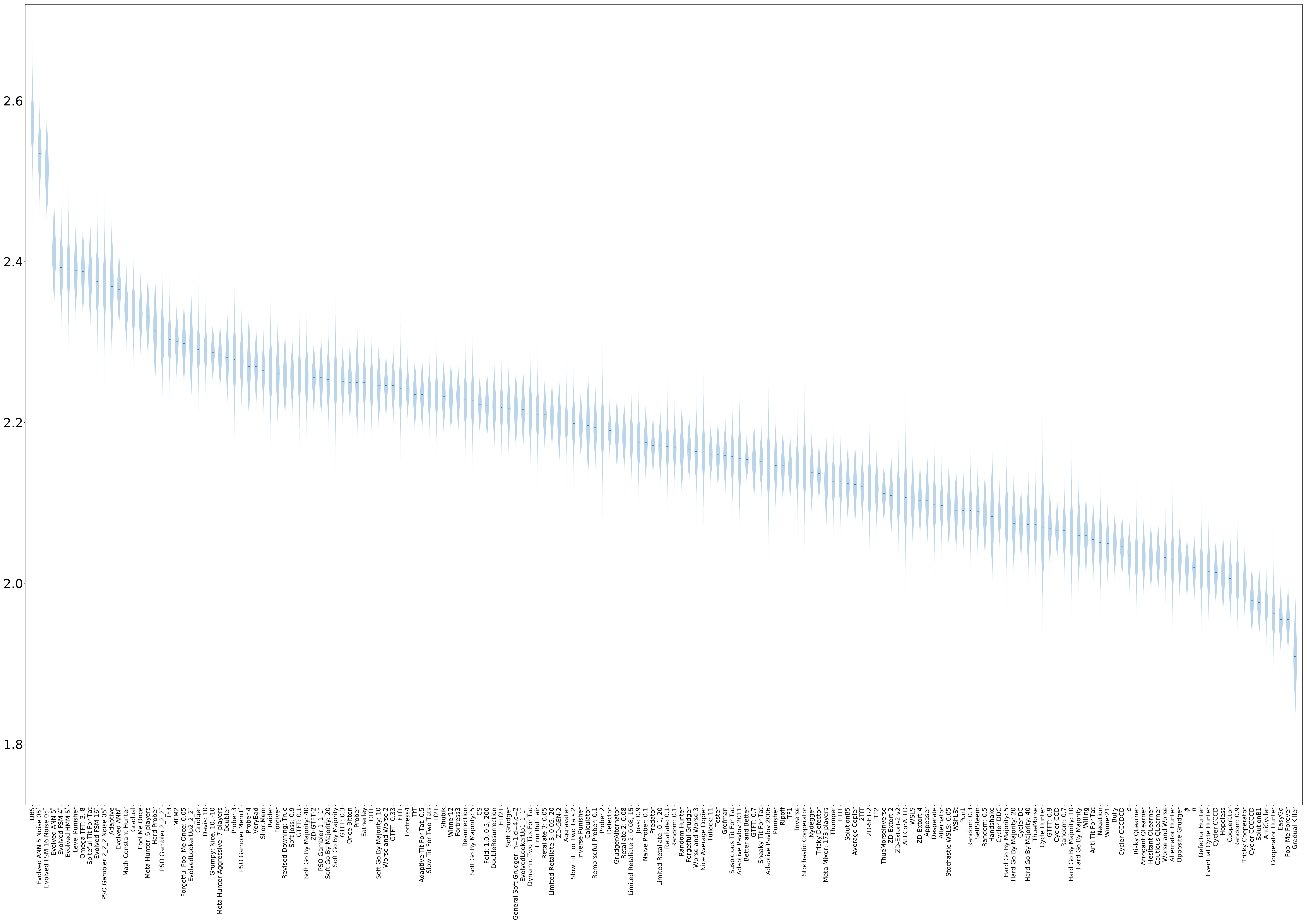}
        \caption{Noisy (5\%) Tournament: Mean score per turn (ranked by median
        over
        \protect\input{./assets/noisy_number_of_repetitions.tex}tournaments).}
        \label{fig:noisy_score}
    \end{figure}
\end{landscape}

Noisy tournaments in which there is a 5\% chance that an action is
flipped are now described. As shown in Table~\ref{tbl:noisy_score} and
Figure~\ref{fig:noisy_score}, the best performing strategies in median payoff
are DBS, designed to account for noise, followed by two strategies trained in
the presence of noise and three trained strategies trained without noise. One of
the strategies trained with noise (PSO Gambler) actually performs less well than
some of the other high ranking strategies including
Spiteful TFT (TFT but defects indefinitely if the opponent defects twice
consecutively) and OmegaTFT (also designed to handle noise). While DBS is the clear
winner, it comes at a 6x increased run time over Evolved FSM 16 Noise 05.

\begin{table}[!hbtp]
    \centering
        \begin{tabular}{lrrrrrrrrr}
\toprule
{} &   mean &    std &    min &     5\% &    25\% &    50\% &    75\% &    95\% &    max \\
\midrule
DBS                              &  2.573 &  0.025 &  2.474 &  2.533 &  2.556 &  2.573 &  2.589 &  2.614 &  2.675 \\
Evolved ANN 5 Noise 05$^{*}$     &  2.534 &  0.025 &  2.418 &  2.492 &  2.517 &  2.534 &  2.551 &  2.575 &  2.629 \\
Evolved FSM 16 Noise 05$^{*}$    &  2.515 &  0.031 &  2.374 &  2.464 &  2.494 &  2.515 &  2.536 &  2.565 &  2.642 \\
Evolved ANN 5$^{*}$              &  2.410 &  0.030 &  2.273 &  2.359 &  2.389 &  2.410 &  2.430 &  2.459 &  2.536 \\
Evolved FSM 4$^{*}$              &  2.393 &  0.027 &  2.286 &  2.348 &  2.374 &  2.393 &  2.411 &  2.437 &  2.505 \\
Evolved HMM 5$^{*}$              &  2.392 &  0.026 &  2.289 &  2.348 &  2.374 &  2.392 &  2.409 &  2.435 &  2.493 \\
Level Punisher                   &  2.388 &  0.025 &  2.281 &  2.347 &  2.372 &  2.389 &  2.405 &  2.429 &  2.503 \\
Omega TFT: 3, 8                  &  2.387 &  0.026 &  2.270 &  2.344 &  2.370 &  2.388 &  2.405 &  2.430 &  2.498 \\
Spiteful Tit For Tat             &  2.383 &  0.030 &  2.259 &  2.334 &  2.363 &  2.383 &  2.403 &  2.432 &  2.517 \\
Evolved FSM 16$^{*}$             &  2.375 &  0.029 &  2.239 &  2.326 &  2.355 &  2.375 &  2.395 &  2.423 &  2.507 \\
PSO Gambler 2\_2\_2 Noise 05$^{*}$ &  2.371 &  0.029 &  2.250 &  2.323 &  2.352 &  2.371 &  2.390 &  2.418 &  2.480 \\
Adaptive                         &  2.369 &  0.038 &  2.217 &  2.306 &  2.344 &  2.369 &  2.395 &  2.431 &  2.524 \\
Evolved ANN$^{*}$                &  2.365 &  0.022 &  2.270 &  2.329 &  2.351 &  2.366 &  2.380 &  2.401 &  2.483 \\
Math Constant Hunter             &  2.344 &  0.022 &  2.257 &  2.308 &  2.329 &  2.344 &  2.359 &  2.382 &  2.445 \\
Gradual                          &  2.341 &  0.021 &  2.248 &  2.306 &  2.327 &  2.341 &  2.355 &  2.376 &  2.429 \\
\bottomrule
\end{tabular}

        \caption{Noisy (5\%) Tournament: Mean score per turn of top 15 strategies
        (ranked by median over
        \protect 50000tournaments)
        ~$^{*}$ indicates that the strategy was trained.}
        \label{tbl:noisy_score}
\end{table}

\begin{figure}[!hbtp]
    \centering
    \includegraphics[width=\textwidth]{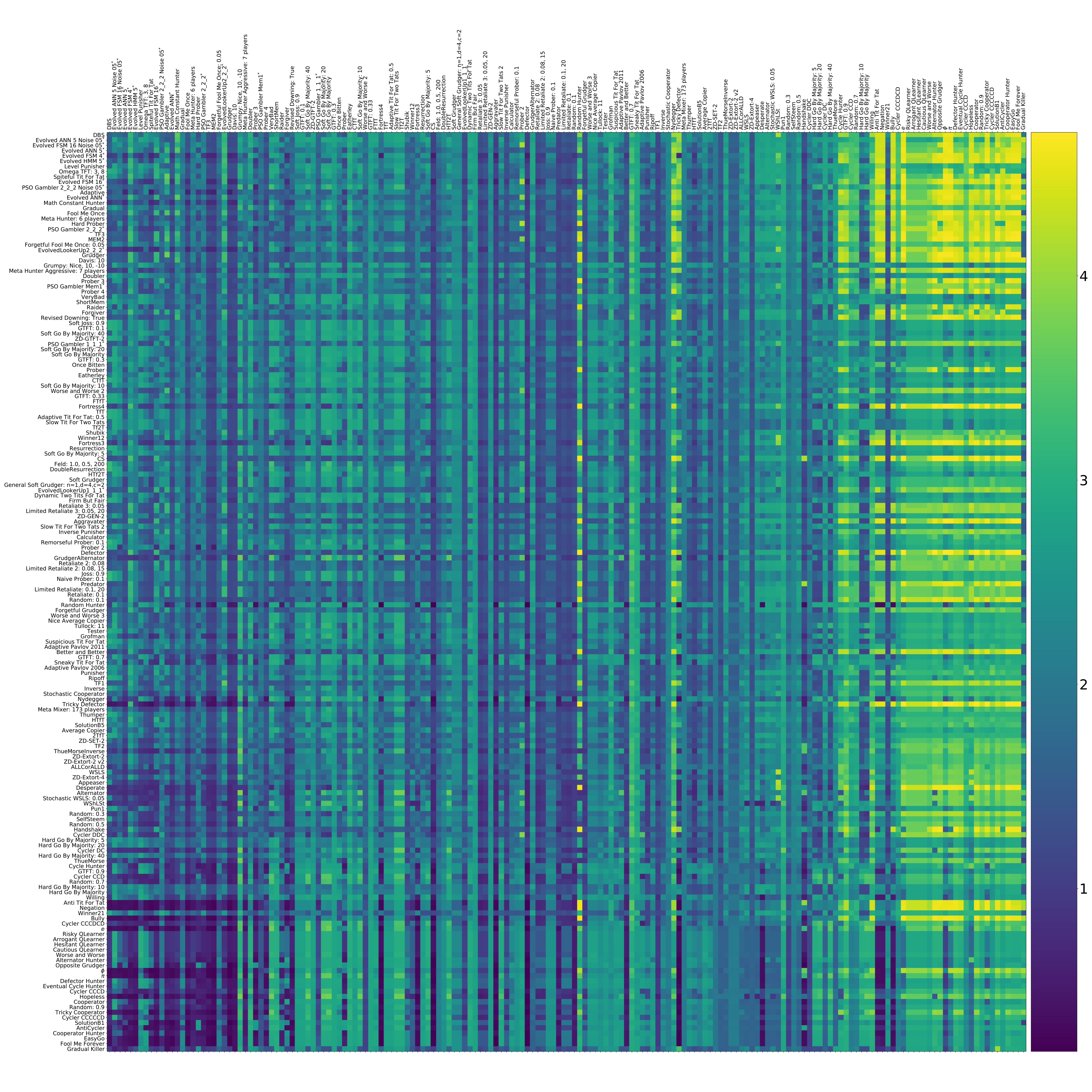}
    \caption{Noisy (5\%) Tournament: Mean score per turn of row players against
    column players (ranked by median over
        \protect\input{./assets/noisy_number_of_repetitions.tex}tournaments).}
    \label{fig:noisy_heatmap}
\end{figure}

Recalling Table~\ref{tbl:standard_score}, the strategies trained in the presence
of noise are also among the best performers in the absence of noise. As shown in
Figure~\ref{fig:noisy_heatmap} the cluster of mutually cooperative strategies is
broken by the noise at 5\%. A similar collection of players excels at winning
matches but again they have a poor total payoff.

\begin{landscape}
    \begin{figure}[!hbtp]
        \centering
        \includegraphics[width=\paperwidth]{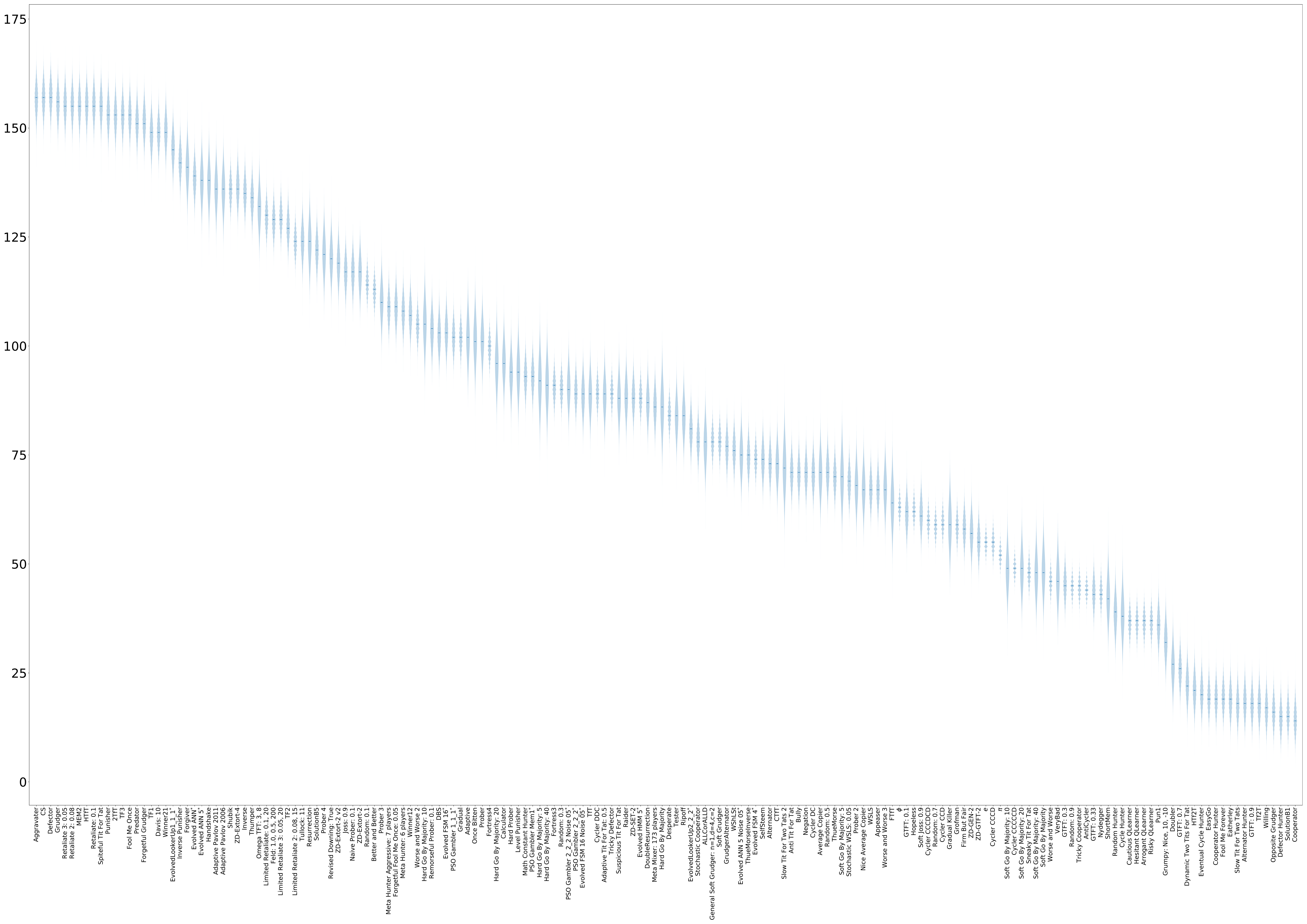}
        \caption{Noisy (5\%) Tournament: number of wins per tournament (ranked by
        median over
        \protect\input{./assets/noisy_number_of_repetitions.tex}tournaments).}
        \label{fig:noisy_winplot}
    \end{figure}
\end{landscape}

As shown in Table~\ref{tbl:noisy_wins_top_winners} and
Figure~\ref{fig:noisy_winplot} the strategies tallying the most wins are
somewhat similar to the standard tournaments, with Defector, the handshaking
CollectiveStrategy~\cite{Li2009}, and Aggravate appearing as the top three again.

\begin{table}[!hbtp]
    \centering
        \begin{tabular}{lrrrrrrrrr}
\toprule
{} &     mean &    std &  min &     5\% &    25\% &    50\% &    75\% &    95\% &  max \\
\midrule
Aggravater           &  156.654 &  3.328 &  141 &  151.0 &  154.0 &  157.0 &  159.0 &  162.0 &  170 \\
CS                   &  156.875 &  3.265 &  144 &  151.0 &  155.0 &  157.0 &  159.0 &  162.0 &  169 \\
Defector             &  157.324 &  3.262 &  144 &  152.0 &  155.0 &  157.0 &  160.0 &  163.0 &  170 \\
Grudger              &  155.590 &  3.303 &  143 &  150.0 &  153.0 &  156.0 &  158.0 &  161.0 &  168 \\
Retaliate 3: 0.05    &  155.382 &  3.306 &  141 &  150.0 &  153.0 &  155.0 &  158.0 &  161.0 &  169 \\
Retaliate 2: 0.08    &  155.365 &  3.320 &  140 &  150.0 &  153.0 &  155.0 &  158.0 &  161.0 &  169 \\
MEM2                 &  155.052 &  3.349 &  140 &  149.0 &  153.0 &  155.0 &  157.0 &  160.0 &  169 \\
HTfT                 &  155.298 &  3.344 &  141 &  150.0 &  153.0 &  155.0 &  158.0 &  161.0 &  168 \\
Retaliate: 0.1       &  155.370 &  3.314 &  139 &  150.0 &  153.0 &  155.0 &  158.0 &  161.0 &  168 \\
Spiteful Tit For Tat &  155.030 &  3.326 &  133 &  150.0 &  153.0 &  155.0 &  157.0 &  160.0 &  167 \\
Punisher             &  153.281 &  3.375 &  140 &  148.0 &  151.0 &  153.0 &  156.0 &  159.0 &  167 \\
2TfT                 &  152.823 &  3.429 &  138 &  147.0 &  151.0 &  153.0 &  155.0 &  158.0 &  165 \\
TF3                  &  153.031 &  3.327 &  138 &  148.0 &  151.0 &  153.0 &  155.0 &  158.0 &  166 \\
Fool Me Once         &  152.817 &  3.344 &  138 &  147.0 &  151.0 &  153.0 &  155.0 &  158.0 &  166 \\
Predator             &  151.406 &  3.403 &  138 &  146.0 &  149.0 &  151.0 &  154.0 &  157.0 &  165 \\
\bottomrule
\end{tabular}

        \caption{Noisy (5\%) Tournament: Number of wins per tournament
        of top 15 strategies (ranked by median wins over
        \protect 50000tournaments).}
        \label{tbl:noisy_wins_top_winners}
\end{table}

As shown in Table~\ref{tbl:noisy_wins}, the top ranking strategies win a larger
number of matches in the presence of noise. For example Spiteful Tit For Tat~\cite{Prison1998}
in one tournament won almost all its matches (167).

\begin{table}[!hbtp]
    \centering
        \begin{tabular}{lrrrrrrrrr}
\toprule
{} &     mean &    std &  min &     5\% &    25\% &    50\% &    75\% &    95\% &  max \\
\midrule
DBS                              &  102.545 &  3.671 &   87 &   97.0 &  100.0 &  103.0 &  105.0 &  109.0 &  118 \\
Evolved ANN 5 Noise 05$^{*}$     &   75.026 &  4.226 &   57 &   68.0 &   72.0 &   75.0 &   78.0 &   82.0 &   93 \\
Evolved FSM 16 Noise 05$^{*}$    &   88.699 &  3.864 &   74 &   82.0 &   86.0 &   89.0 &   91.0 &   95.0 &  104 \\
Evolved ANN 5$^{*}$              &  137.878 &  4.350 &  118 &  131.0 &  135.0 &  138.0 &  141.0 &  145.0 &  156 \\
Evolved FSM 4$^{*}$              &   74.250 &  2.694 &   64 &   70.0 &   72.0 &   74.0 &   76.0 &   79.0 &   85 \\
Evolved HMM 5$^{*}$              &   88.189 &  2.774 &   77 &   84.0 &   86.0 &   88.0 &   90.0 &   93.0 &   99 \\
Level Punisher                   &   94.263 &  4.789 &   75 &   86.0 &   91.0 &   94.0 &   97.0 &  102.0 &  116 \\
Omega TFT: 3, 8                  &  131.655 &  4.302 &  112 &  125.0 &  129.0 &  132.0 &  135.0 &  139.0 &  150 \\
Spiteful Tit For Tat             &  155.030 &  3.326 &  133 &  150.0 &  153.0 &  155.0 &  157.0 &  160.0 &  167 \\
Evolved FSM 16$^{*}$             &  103.288 &  3.631 &   89 &   97.0 &  101.0 &  103.0 &  106.0 &  109.0 &  118 \\
PSO Gambler 2\_2\_2 Noise 05$^{*}$ &   90.515 &  4.012 &   75 &   84.0 &   88.0 &   90.0 &   93.0 &   97.0 &  109 \\
Adaptive                         &  101.898 &  4.899 &   83 &   94.0 &   99.0 &  102.0 &  105.0 &  110.0 &  124 \\
Evolved ANN$^{*}$                &  138.514 &  3.401 &  125 &  133.0 &  136.0 &  139.0 &  141.0 &  144.0 &  153 \\
Math Constant Hunter             &   93.010 &  3.254 &   79 &   88.0 &   91.0 &   93.0 &   95.0 &   98.0 &  107 \\
Gradual                          &  101.899 &  2.870 &   91 &   97.0 &  100.0 &  102.0 &  104.0 &  107.0 &  114 \\
\bottomrule
\end{tabular}

        \caption{Noisy (5\%) Tournament: Number of wins per tournament
        of top 15 strategies (ranked by median score over
        \protect 50000tournaments).}
        \label{tbl:noisy_wins}
\end{table}

\begin{landscape}
    \begin{figure}[!hbtp]
        \centering
        \includegraphics[width=\paperwidth]{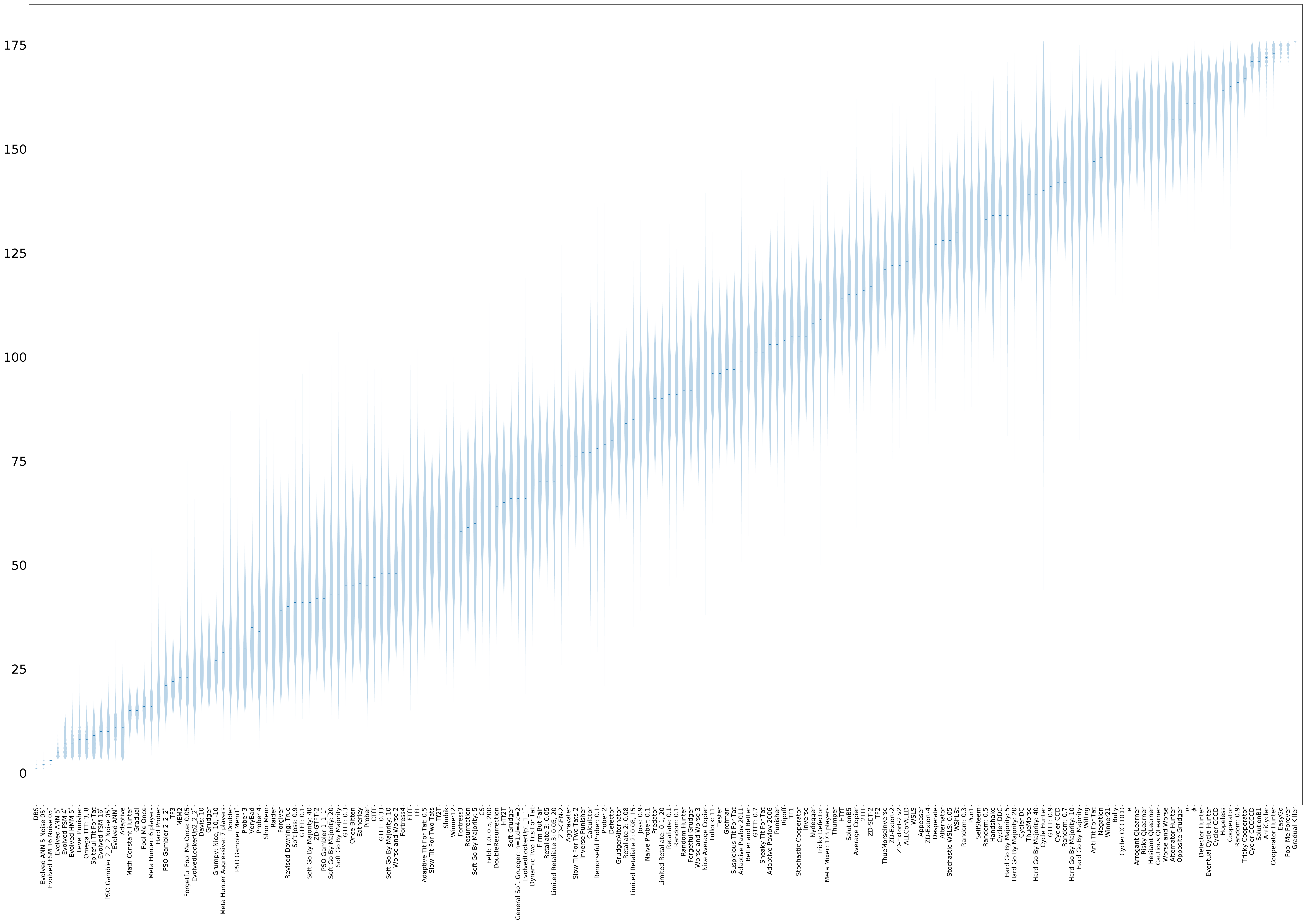}
        \caption{Noisy (5\%) Tournament: rank in each tournament (ranked by
        median over
        \protect\input{./assets/noisy_number_of_repetitions.tex}tournaments).}
        \label{fig:noisy_ranks_boxplot}
    \end{figure}
\end{landscape}

Finally, Table~\ref{tbl:noisy_ranks} and
Figure~\ref{fig:noisy_ranks_boxplot} show the ranks (based on median score)
of each strategy over the repeated tournaments. We see that the stochasticity
of the ranks understandably increases relative to the standard tournament. An
exception is the top three strategies, for example, the DBS strategy never ranks
lower than
second and wins 75\% of the time. The two strategies trained for noisy
tournaments rank in the top three 95\% of the time.

\begin{table}[!hbtp]
    \centering
        \begin{tabular}{lrrrrrrrrr}
\toprule
{} &    mean &    std &  min &      5\% &   25\% &   50\% &   75\% &   95\% &  max \\
\midrule
DBS                              &   1.205 &  0.468 &    1 &   1.000 &   1.0 &   1.0 &   1.0 &   2.0 &    3 \\
Evolved ANN 5 Noise 05$^{*}$     &   2.184 &  0.629 &    1 &   1.000 &   2.0 &   2.0 &   3.0 &   3.0 &    5 \\
Evolved FSM 16 Noise 05$^{*}$    &   2.626 &  0.618 &    1 &   1.000 &   2.0 &   3.0 &   3.0 &   3.0 &    9 \\
Evolved ANN 5$^{*}$              &   6.371 &  2.786 &    2 &   4.000 &   4.0 &   5.0 &   8.0 &  12.0 &   31 \\
Evolved FSM 4$^{*}$              &   7.919 &  3.175 &    3 &   4.000 &   5.0 &   7.0 &  10.0 &  14.0 &   33 \\
Evolved HMM 5$^{*}$              &   7.996 &  3.110 &    3 &   4.000 &   6.0 &   7.0 &  10.0 &  14.0 &   26 \\
Level Punisher                   &   8.337 &  3.083 &    3 &   4.000 &   6.0 &   8.0 &  10.0 &  14.0 &   26 \\
Omega TFT: 3, 8                  &   8.510 &  3.249 &    3 &   4.000 &   6.0 &   8.0 &  11.0 &  14.0 &   32 \\
Spiteful Tit For Tat             &   9.159 &  3.772 &    3 &   4.000 &   6.0 &   9.0 &  12.0 &  16.0 &   40 \\
Evolved FSM 16$^{*}$             &  10.218 &  4.099 &    3 &   4.975 &   7.0 &  10.0 &  13.0 &  17.0 &   56 \\
PSO Gambler 2\_2\_2 Noise 05$^{*}$ &  10.760 &  4.102 &    3 &   5.000 &   8.0 &  10.0 &  13.0 &  18.0 &   47 \\
Evolved ANN$^{*}$                &  11.346 &  3.252 &    3 &   6.000 &   9.0 &  11.0 &  13.0 &  17.0 &   32 \\
Adaptive                         &  11.420 &  5.739 &    3 &   4.000 &   7.0 &  11.0 &  14.0 &  21.0 &   63 \\
Math Constant Hunter             &  14.668 &  3.788 &    3 &   9.000 &  12.0 &  15.0 &  17.0 &  21.0 &   43 \\
Gradual                          &  15.163 &  3.672 &    4 &  10.000 &  13.0 &  15.0 &  17.0 &  21.0 &   49 \\
\bottomrule
\end{tabular}

        \caption{Noisy (5\%) Tournament: Rank in each tournament
        of top 15 strategies (ranked by median over
        \protect 50000tournaments).}
        \label{tbl:noisy_ranks}
\end{table}

\begin{figure}[!hbtp]
    \centering
    \begin{subfigure}[t]{.3\textwidth}
        \centering
        \includegraphics[width=\textwidth]{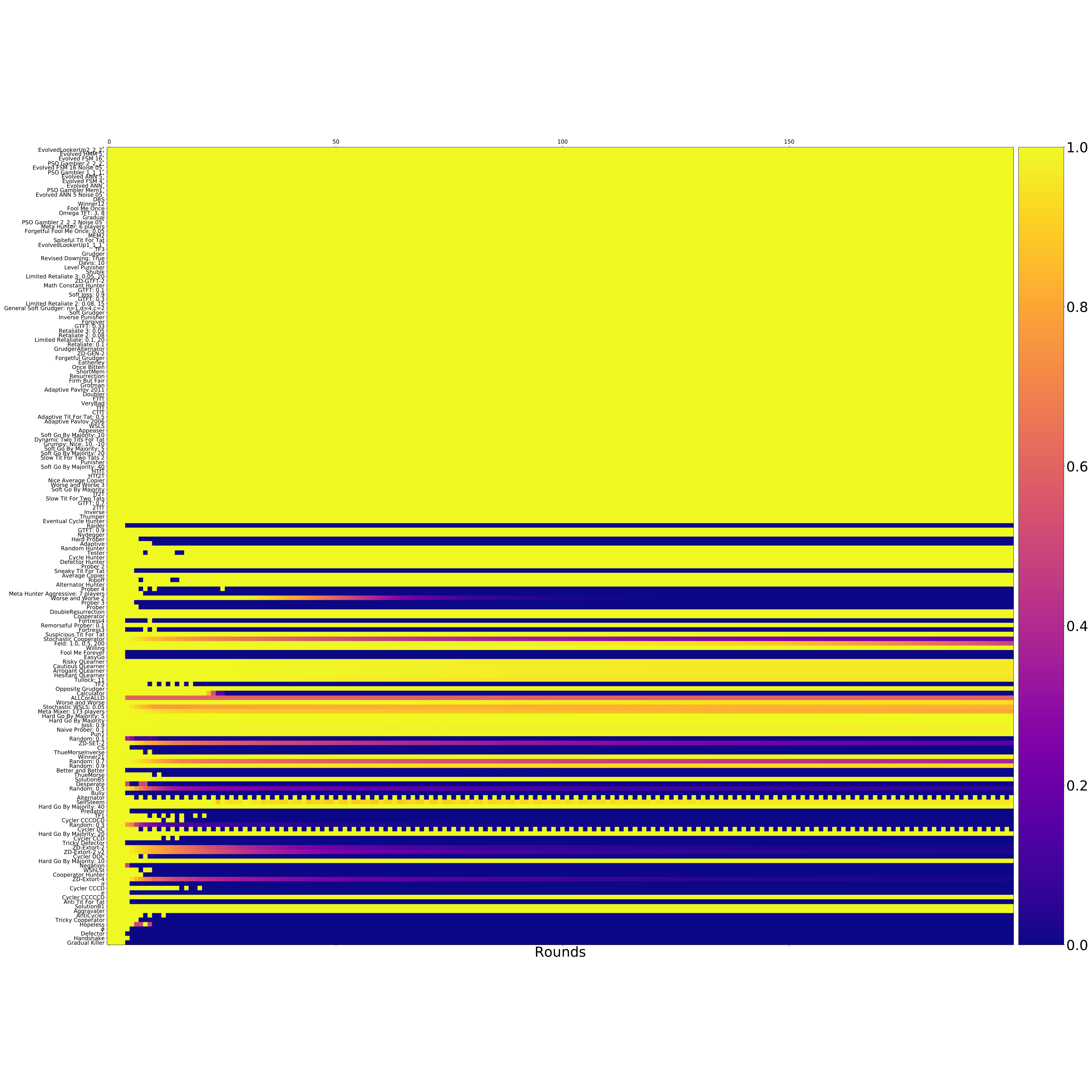}
        \caption{DBS}
    \end{subfigure}%
    ~
    \begin{subfigure}[t]{.3\textwidth}
        \centering
        \includegraphics[width=\textwidth]{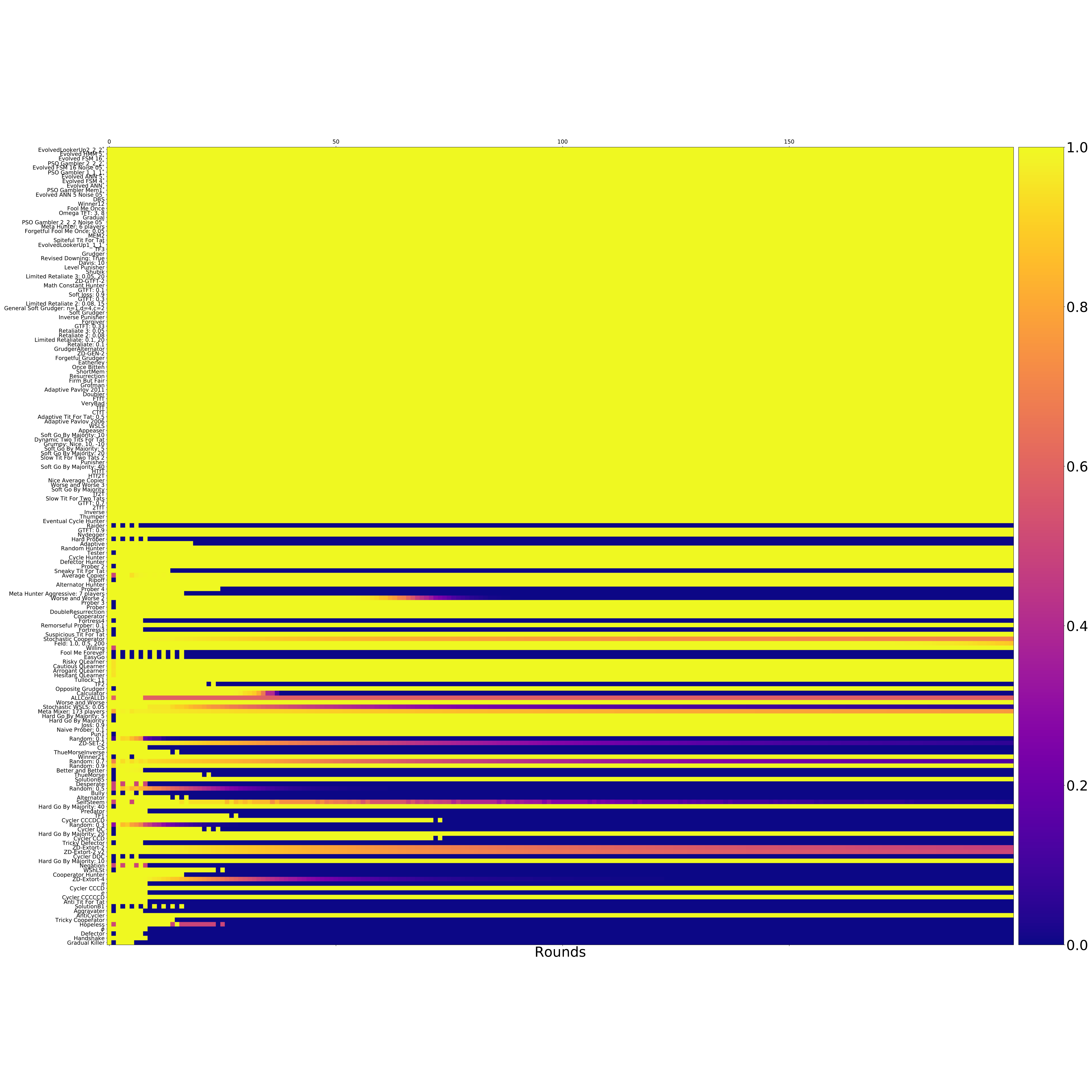}
        \caption{Evolved\_ANN\_5\_Noise\_05}
    \end{subfigure}%
    ~
    \begin{subfigure}[t]{.3\textwidth}
        \centering
        \includegraphics[width=\textwidth]{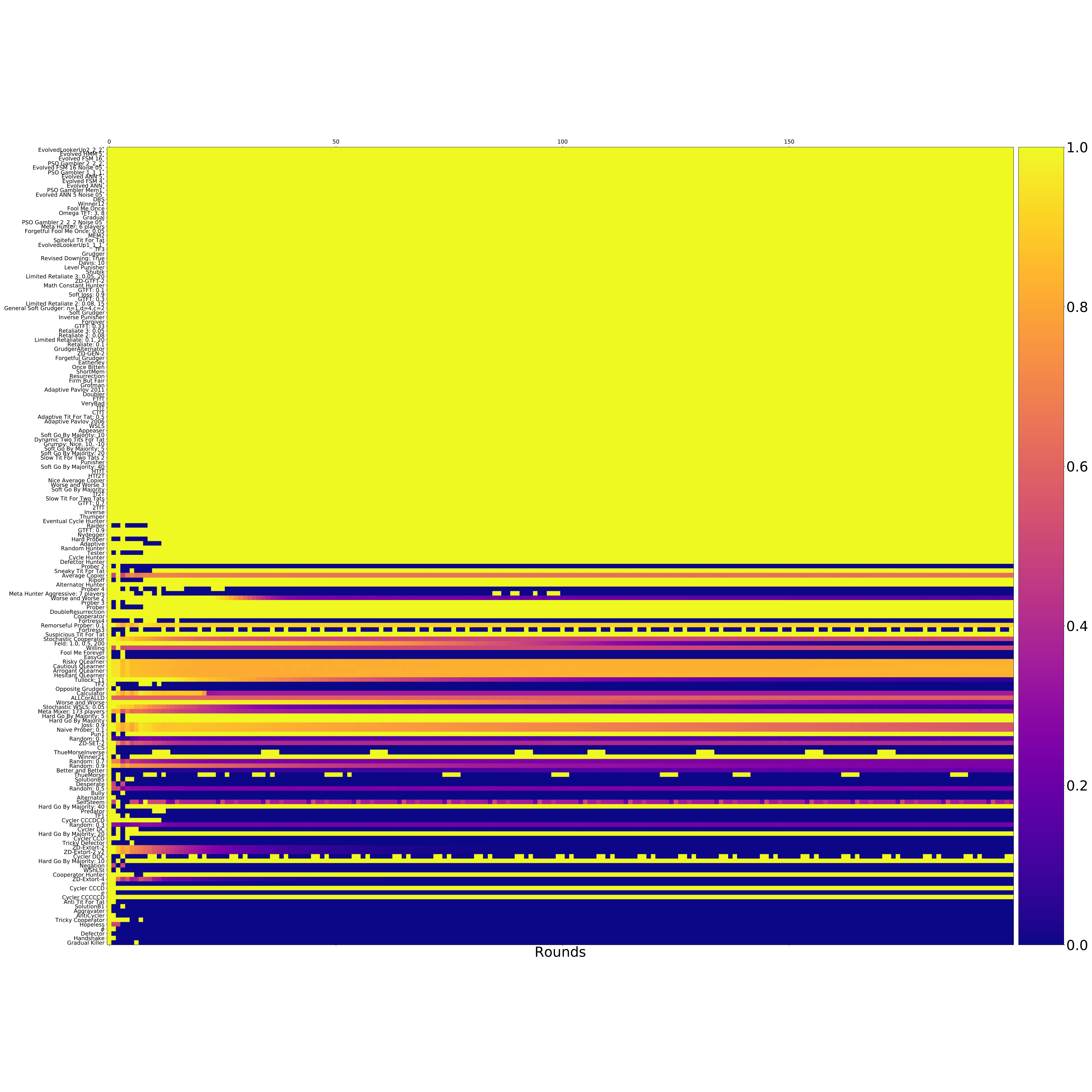}
        \caption{Evolved\_FSM\_16\_Noise\_05}
    \end{subfigure}%

    \caption{Comparison of cooperation rates for the Noisy (5\%) Tournament Top 3
    (over 10000 standard tournaments).}
    \label{fig:comparison_cooperation_heatmaps_noisy}
\end{figure}

Figure~\ref{fig:comparison_cooperation_heatmaps_noisy} shows the rate of
cooperation in each round for the top three strategies (in the absense of noise)
and just as for the top performing strategies in the standard tournament
(Figure~\ref{fig:comparison_cooperation_heatmaps_standard}) it is evident that
the strategies never defect first and learn to quickly punish poorer strategies.

\section{Methods}\label{sec:methods}

The trained strategies (denoted by a \(^{*}\) in
Appendix~\ref{app:list_of_players}) were trained using reinforcement
learning algorithms. The ideas of reinforcement learning can be attributed to
the original work of \cite{turing1950computing} in which the notion that
computers would learn by taking random actions but according to a distribution
that picked actions with high rewards more often. The two particular algorithms
used here:

\begin{itemize}
    \item Particle Swarm Algorithm: \cite{imran2013overview}.
    \item Evolutionary algorithm: \cite{moriarty1999evolutionary}.
\end{itemize}

The Particle Swarm Algorithm is implemented using the pyswarm library:
\url{https://pypi.python.org/pypi/pyswarm}. This algorithm was used only to
train the Gambler archetype.

All other strategies were trained using evolutionary algorithms. The
evolutionary algorithms used standard techniques, varying strategies by
mutation and crossover, and evaluating the performance against each opponent
for many repetitions. The best performing strategies in each generation are
persisted, variants created, and objective functions computed again.

The default parameters for this procedure:

\begin{itemize}
	\item A population size of 40 individuals (kept constant across the
        generations);
	\item A mutation rate of 10\%;
	\item 10 individuals kept from one generation to the next;
    \item A total of 500 generations.
\end{itemize}

All implementations of these algorithms are archived at
\cite{dojo}. This software is (similarly to the Axelrod
library) available on github
\url{https://github.com/Axelrod-Python/axelrod-dojo}. There are objective
functions for:

\begin{itemize}
 \item total or mean payoff,
 \item total or mean payoff difference (unused in this work),
 \item total Moran process wins (fixation probability). This lead to the
     strategies named TF1, TF2, TF3 listed in
     Appendix~\ref{app:list_of_players}.
\end{itemize}

These can be used in noisy or standard environments (as evidenced by
Sections~\ref{sec:standard} and~\ref{sec:noise}). These objectives can be
further modified to suit other purposes. New strategies could be trained with
variations including spatial structure and probabilistically ending matches.

\section{Discussion}

The tournament results indicate that pre-trained strategies are generally better
than human designed strategies at maximizing payoff against a diverse set of
opponents. An evolutionary algorithm produces strategies based on multiple
generic archetypes that are able to achieve a higher average
score than any other known opponent in a standard tournament. Most of the trained
strategies use multiple rounds of the history of play (some using all of it) and
outperform memory-one strategies from the literature. Interestingly, a trained
memory one strategy produced by a particle swarm algorithm performs well, better
than human designed strategies such as Win Stay Lose Shift and zero determinant
strategies (which enforce a payoff difference rather than maximize total payoff).
The generic structure of the trained strategies did not appear to be
critical for the standard tournament -- strategies based on lookup tables,
finite state machines, neural networks, and stochastic variants all performed well.
Single layer neural networks (Section~\ref{sec:ann}) performed well in both
noisy and standard tournaments
though these had some aspect of human involvement in the selection of features.
This is in line with the other strategies also where some human decisions are made
regarding the structure. For the LookerUp and Gambler archetypes
(Sections~\ref{sec:lookerup} and~\ref{sec:gambler})
a decision has to be made
regarding the number of rounds of history and initial play that are to be used.
In contrast, the finite state machines and hidden Markov models
(Sections\ref{sec:fsm} and~\ref{sec:hmm}) required only a choice of the number
of states, and the training algorithm can eliminate unneeded states in the case
of finite state machines (evidenced by the unconnected nodes in the diagrams
for the included representations).

Many strategies can be represented by multiple archetypes, however some
archetypes will be more efficient in encoding the patterns present in the data.
The fact that the Lookerup strategy does the best for the standard
tournament indicates that it represents an efficient reduction of
dimension which in turn makes its training more efficient. In particular the
first rounds of play were valuable bits of information. For the noisy
tournament however the dimension reduction represented by some archetypes
indicates that some features of the data are not captured by the lookup
tables while they are by the neural networks and the finite state machines,
allowing the latter to adapt better to the noisy environment. Intuitively, a noisy
environment can significantly affect a lookup table based on the last two rounds
of play since these action pairs compete with probing defections, apologies, and
retaliations. Accordingly, it is not surprising that additional parameter space
is needed to adapt to a noisy environment.

In opposition to historical tournament results and community folklore,
our results show that complex strategies can be very effective for the
IPD\@. Designing complex strategies for the prisoner's dilemma appears to be
difficult for humans. Of all the human-designed
strategies in the library, only DBS consistently performs well, and it is
substantially more complex than traditional tournament winners like TFT, OmegaTFT,
and zero determinant strategies. Furthermore, dealing with noise is difficult
for most strategies. Two strategies designed specifically to account for noise,
DBS and OmegaTFT, perform well and only DBS performs better than the trained
strategies and \textbf{only} in some noisy contexts. Empirically we find that
DBS (with its default parameters) does not win tournaments at 1\% noise.
However DBS has a parameter that accounts for the expected amount of noise and a
followup study with various noise levels could make a more complete study of
the performance of DBS and strategies trained at various noise levels.

The strategies trained to maximize their average score are generally
cooperative and do not defect first. Maximizing for individual
performance across a collection of opponents leads to mutual cooperation despite
the fact that mutual cooperation is an unstable evolutionary equilibrium for the prisoner's
dilemma. Specifically it is noted that the reinforcement learning process for maximizing
payoff does not lead to exploitative zero determinant strategies, which may also
be a result of the collection of training strategies, of which several retaliate
harshly. Training with the objective of maximizing payoff difference may
produce strategies more like zero determinant strategies.

For the trained
strategies utilizing look up tables we generally found those that incorporate
one or more of the initial rounds of play outperformed those that did not. The
strategies based on neural networks and finite state machines also are able to
condition throughout a match on the first rounds of play. Accordingly, we conclude
that first impressions matter in the IPD\@. The best strategies are nice (never
defecting first) and the impact of the first rounds of play could be further
investigated with the Axelrod library
in future work by e.g.\ forcing all strategies to defect on the first round.

Finally, we note that as the library grows, the top performing strategies
sometimes shuffle, and are not retrained automatically. Most of the strategies were
trained on an earlier version of the library (v2.2.0: \cite{axelrodproject2.2})
that did not include DBS and several other opponents. The precise parameters
that are optimal will depend on the pool of opponents. Moreover we have not
extensively trained strategies to determine the minimum parameter spaces that are
sufficient -- neural networks with fewer nodes and features and finite state
machines with fewer states may suffice. See \cite{ashlock2013impact} for
discussion of resource availability for IPD strategies.


\section*{Acknowledgements}

This work was performed using the computational facilities of the Advanced
Research Computing @ Cardiff (ARCCA) Division, Cardiff University.

A variety of software libraries have been used in this work:

\begin{itemize}
    \item The Axelrod library (IPD strategies and Tournaments)
        \cite{axelrodproject}.
    \item The matplotlib library (visualisation) \cite{hunter2007matplotlib}.
    \item The pandas and numpy libraries (data manipulation)
        \cite{mckinney2010data, walt2011numpy}.
\end{itemize}

\bibliographystyle{plain}
\bibliography{bibliography.bib}

\appendix

\section{List of players}\label{app:list_of_players}

The players used for this study are from Axelrod version 2.13.0
\cite{axelrodproject}.

\begin{multicols}{2}
	\begin{enumerate}
		\item $\phi$ - \textit{Deterministic} - \textit{Memory depth}: \(\infty\). \cite{axelrodproject}
\item $\pi$ - \textit{Deterministic} - \textit{Memory depth}: \(\infty\). \cite{axelrodproject}
\item $e$ - \textit{Deterministic} - \textit{Memory depth}: \(\infty\). \cite{axelrodproject}
\item ALLCorALLD - \textit{Stochastic} - \textit{Memory depth}: 1. \cite{axelrodproject}
\item Adaptive - \textit{Deterministic} - \textit{Memory depth}: \(\infty\). \cite{Li2011}
\item Adaptive Pavlov 2006 - \textit{Deterministic} - \textit{Memory depth}: \(\infty\). \cite{kendall2007iterated}
\item Adaptive Pavlov 2011 - \textit{Deterministic} - \textit{Memory depth}: \(\infty\). \cite{Li2011}
\item Adaptive Tit For Tat: 0.5 - \textit{Deterministic} - \textit{Memory depth}: \(\infty\). \cite{Tzafestas2000}
\item Aggravater - \textit{Deterministic} - \textit{Memory depth}: \(\infty\). \cite{axelrodproject}
\item Alternator - \textit{Deterministic} - \textit{Memory depth}: 1. \cite{Axelrod1984, Mittal2009}
\item Alternator Hunter - \textit{Deterministic} - \textit{Memory depth}: \(\infty\). \cite{axelrodproject}
\item Anti Tit For Tat - \textit{Deterministic} - \textit{Memory depth}: 1. \cite{Hilbe2013}
\item AntiCycler - \textit{Deterministic} - \textit{Memory depth}: \(\infty\). \cite{axelrodproject}
\item Appeaser - \textit{Deterministic} - \textit{Memory depth}: \(\infty\). \cite{axelrodproject}
\item Arrogant QLearner - \textit{Stochastic} - \textit{Memory depth}: \(\infty\). \cite{axelrodproject}
\item Average Copier - \textit{Stochastic} - \textit{Memory depth}: \(\infty\). \cite{axelrodproject}
\item Better and Better - \textit{Stochastic} - \textit{Memory depth}: \(\infty\). \cite{Prison1998}
\item Bully - \textit{Deterministic} - \textit{Memory depth}: 1. \cite{Nachbar1992}
\item Calculator - \textit{Stochastic} - \textit{Memory depth}: \(\infty\). \cite{Prison1998}
\item Cautious QLearner - \textit{Stochastic} - \textit{Memory depth}: \(\infty\). \cite{axelrodproject}
\item CollectiveStrategy (\textbf{CS}) - \textit{Deterministic} - \textit{Memory depth}: \(\infty\). \cite{Li2009}
\item Contrite Tit For Tat (\textbf{CTfT}) - \textit{Deterministic} - \textit{Memory depth}: 3. \cite{Axelrod1995}
\item Cooperator - \textit{Deterministic} - \textit{Memory depth}: 0. \cite{Axelrod1984, Mittal2009, Press2012}
\item Cooperator Hunter - \textit{Deterministic} - \textit{Memory depth}: \(\infty\). \cite{axelrodproject}
\item Cycle Hunter - \textit{Deterministic} - \textit{Memory depth}: \(\infty\). \cite{axelrodproject}
\item Cycler CCCCCD - \textit{Deterministic} - \textit{Memory depth}: 5. \cite{axelrodproject}
\item Cycler CCCD - \textit{Deterministic} - \textit{Memory depth}: 3. \cite{axelrodproject}
\item Cycler CCCDCD - \textit{Deterministic} - \textit{Memory depth}: 5. \cite{axelrodproject}
\item Cycler CCD - \textit{Deterministic} - \textit{Memory depth}: 2. \cite{Mittal2009}
\item Cycler DC - \textit{Deterministic} - \textit{Memory depth}: 1. \cite{axelrodproject}
\item Cycler DDC - \textit{Deterministic} - \textit{Memory depth}: 2. \cite{Mittal2009}
\item DBS: 0.75, 3, 4, 3, 5 - \textit{Deterministic} - \textit{Memory depth}: \(\infty\). \cite{Au2006}
\item Davis: 10 - \textit{Deterministic} - \textit{Memory depth}: \(\infty\). \cite{Axelrod1980}
\item Defector - \textit{Deterministic} - \textit{Memory depth}: 0. \cite{Axelrod1984, Mittal2009, Press2012}
\item Defector Hunter - \textit{Deterministic} - \textit{Memory depth}: \(\infty\). \cite{axelrodproject}
\item Desperate - \textit{Stochastic} - \textit{Memory depth}: 1. \cite{Berg2015}
\item DoubleResurrection - \textit{Deterministic} - \textit{Memory depth}: 5. \cite{Eckhart2015}
\item Doubler - \textit{Deterministic} - \textit{Memory depth}: \(\infty\). \cite{Prison1998}
\item Dynamic Two Tits For Tat - \textit{Stochastic} - \textit{Memory depth}: 2. \cite{axelrodproject}
\item EasyGo - \textit{Deterministic} - \textit{Memory depth}: \(\infty\). \cite{Li2011, Prison1998}
\item Eatherley - \textit{Stochastic} - \textit{Memory depth}: \(\infty\). \cite{Axelrod1980b}
\item Eventual Cycle Hunter - \textit{Deterministic} - \textit{Memory depth}: \(\infty\). \cite{axelrodproject}
\item Evolved ANN - \textit{Deterministic} - \textit{Memory depth}: \(\infty\). \cite{axelrodproject}
\item Evolved ANN 5 - \textit{Deterministic} - \textit{Memory depth}: \(\infty\). \cite{axelrodproject}
\item Evolved ANN 5 Noise 05 - \textit{Deterministic} - \textit{Memory depth}: \(\infty\). \cite{axelrodproject}
\item Evolved FSM 16 - \textit{Deterministic} - \textit{Memory depth}: 16. \cite{axelrodproject}
\item Evolved FSM 16 Noise 05 - \textit{Deterministic} - \textit{Memory depth}: 16. \cite{axelrodproject}
\item Evolved FSM 4 - \textit{Deterministic} - \textit{Memory depth}: 4. \cite{axelrodproject}
\item Evolved HMM 5 - \textit{Stochastic} - \textit{Memory depth}: 5. \cite{axelrodproject}
\item EvolvedLookerUp1\_1\_1 - \textit{Deterministic} - \textit{Memory depth}: \(\infty\). \cite{axelrodproject}
\item EvolvedLookerUp2\_2\_2 - \textit{Deterministic} - \textit{Memory depth}: \(\infty\). \cite{axelrodproject}
\item Feld: 1.0, 0.5, 200 - \textit{Stochastic} - \textit{Memory depth}: 200. \cite{Axelrod1980}
\item Firm But Fair - \textit{Stochastic} - \textit{Memory depth}: 1. \cite{Frean1994}
\item Fool Me Forever - \textit{Deterministic} - \textit{Memory depth}: \(\infty\). \cite{axelrodproject}
\item Fool Me Once - \textit{Deterministic} - \textit{Memory depth}: \(\infty\). \cite{axelrodproject}
\item Forgetful Fool Me Once: 0.05 - \textit{Stochastic} - \textit{Memory depth}: \(\infty\). \cite{axelrodproject}
\item Forgetful Grudger - \textit{Deterministic} - \textit{Memory depth}: 10. \cite{axelrodproject}
\item Forgiver - \textit{Deterministic} - \textit{Memory depth}: \(\infty\). \cite{axelrodproject}
\item Forgiving Tit For Tat (\textbf{FTfT}) - \textit{Deterministic} - \textit{Memory depth}: \(\infty\). \cite{axelrodproject}
\item Fortress3 - \textit{Deterministic} - \textit{Memory depth}: 3. \cite{Ashlock2006b}
\item Fortress4 - \textit{Deterministic} - \textit{Memory depth}: 4. \cite{Ashlock2006b}
\item GTFT: 0.1 - \textit{Stochastic} - \textit{Memory depth}: 1.
\item GTFT: 0.3 - \textit{Stochastic} - \textit{Memory depth}: 1.
\item GTFT: 0.33 - \textit{Stochastic} - \textit{Memory depth}: 1. \cite{Gaudesi2016, Nowak1993}
\item GTFT: 0.7 - \textit{Stochastic} - \textit{Memory depth}: 1.
\item GTFT: 0.9 - \textit{Stochastic} - \textit{Memory depth}: 1.
\item General Soft Grudger: n=1,d=4,c=2 - \textit{Deterministic} - \textit{Memory depth}: \(\infty\). \cite{axelrodproject}
\item Gradual - \textit{Deterministic} - \textit{Memory depth}: \(\infty\). \cite{Beaufils1997}
\item Gradual Killer: ('D', 'D', 'D', 'D', 'D', 'C', 'C') - \textit{Deterministic} - \textit{Memory depth}: \(\infty\). \cite{Prison1998}
\item Grofman - \textit{Stochastic} - \textit{Memory depth}: \(\infty\). \cite{Axelrod1980}
\item Grudger - \textit{Deterministic} - \textit{Memory depth}: 1. \cite{Axelrod1980, Banks1990, Beaufils1997, Berg2015, Li2011}
\item GrudgerAlternator - \textit{Deterministic} - \textit{Memory depth}: \(\infty\). \cite{Prison1998}
\item Grumpy: Nice, 10, -10 - \textit{Deterministic} - \textit{Memory depth}: \(\infty\). \cite{axelrodproject}
\item Handshake - \textit{Deterministic} - \textit{Memory depth}: \(\infty\). \cite{Robson1990}
\item Hard Go By Majority - \textit{Deterministic} - \textit{Memory depth}: \(\infty\). \cite{Mittal2009}
\item Hard Go By Majority: 10 - \textit{Deterministic} - \textit{Memory depth}: 10. \cite{axelrodproject}
\item Hard Go By Majority: 20 - \textit{Deterministic} - \textit{Memory depth}: 20. \cite{axelrodproject}
\item Hard Go By Majority: 40 - \textit{Deterministic} - \textit{Memory depth}: 40. \cite{axelrodproject}
\item Hard Go By Majority: 5 - \textit{Deterministic} - \textit{Memory depth}: 5. \cite{axelrodproject}
\item Hard Prober - \textit{Deterministic} - \textit{Memory depth}: \(\infty\). \cite{Prison1998}
\item Hard Tit For 2 Tats (\textbf{HTf2T}) - \textit{Deterministic} - \textit{Memory depth}: 3. \cite{Stewart2012}
\item Hard Tit For Tat (\textbf{HTfT}) - \textit{Deterministic} - \textit{Memory depth}: 3. \cite{PD2017}
\item Hesitant QLearner - \textit{Stochastic} - \textit{Memory depth}: \(\infty\). \cite{axelrodproject}
\item Hopeless - \textit{Stochastic} - \textit{Memory depth}: 1. \cite{Berg2015}
\item Inverse - \textit{Stochastic} - \textit{Memory depth}: \(\infty\). \cite{axelrodproject}
\item Inverse Punisher - \textit{Deterministic} - \textit{Memory depth}: \(\infty\). \cite{axelrodproject}
\item Joss: 0.9 - \textit{Stochastic} - \textit{Memory depth}: 1. \cite{Axelrod1980, Stewart2012}
\item Level Punisher - \textit{Deterministic} - \textit{Memory depth}: \(\infty\). \cite{Eckhart2015}
\item Limited Retaliate 2: 0.08, 15 - \textit{Deterministic} - \textit{Memory depth}: \(\infty\). \cite{axelrodproject}
\item Limited Retaliate 3: 0.05, 20 - \textit{Deterministic} - \textit{Memory depth}: \(\infty\). \cite{axelrodproject}
\item Limited Retaliate: 0.1, 20 - \textit{Deterministic} - \textit{Memory depth}: \(\infty\). \cite{axelrodproject}
\item MEM2 - \textit{Deterministic} - \textit{Memory depth}: \(\infty\). \cite{Li2014}
\item Math Constant Hunter - \textit{Deterministic} - \textit{Memory depth}: \(\infty\). \cite{axelrodproject}
\item Meta Hunter Aggressive: 7 players - \textit{Deterministic} - \textit{Memory depth}: \(\infty\). \cite{axelrodproject}
\item Meta Hunter: 6 players - \textit{Deterministic} - \textit{Memory depth}: \(\infty\). \cite{axelrodproject}
\item Meta Mixer: 173 players - \textit{Stochastic} - \textit{Memory depth}: \(\infty\). \cite{axelrodproject}
\item Naive Prober: 0.1 - \textit{Stochastic} - \textit{Memory depth}: 1. \cite{Li2011}
\item Negation - \textit{Stochastic} - \textit{Memory depth}: 1. \cite{PD2017}
\item Nice Average Copier - \textit{Stochastic} - \textit{Memory depth}: \(\infty\). \cite{axelrodproject}
\item Nydegger - \textit{Deterministic} - \textit{Memory depth}: 3. \cite{Axelrod1980}
\item Omega TFT: 3, 8 - \textit{Deterministic} - \textit{Memory depth}: \(\infty\). \cite{kendall2007iterated}
\item Once Bitten - \textit{Deterministic} - \textit{Memory depth}: 12. \cite{axelrodproject}
\item Opposite Grudger - \textit{Deterministic} - \textit{Memory depth}: \(\infty\). \cite{axelrodproject}
\item PSO Gambler 1\_1\_1 - \textit{Stochastic} - \textit{Memory depth}: \(\infty\). \cite{axelrodproject}
\item PSO Gambler 2\_2\_2 - \textit{Stochastic} - \textit{Memory depth}: \(\infty\). \cite{axelrodproject}
\item PSO Gambler 2\_2\_2 Noise 05 - \textit{Stochastic} - \textit{Memory depth}: \(\infty\). \cite{axelrodproject}
\item PSO Gambler Mem1 - \textit{Stochastic} - \textit{Memory depth}: 1. \cite{axelrodproject}
\item Predator - \textit{Deterministic} - \textit{Memory depth}: 9. \cite{Ashlock2006b}
\item Prober - \textit{Deterministic} - \textit{Memory depth}: \(\infty\). \cite{Li2011}
\item Prober 2 - \textit{Deterministic} - \textit{Memory depth}: \(\infty\). \cite{Prison1998}
\item Prober 3 - \textit{Deterministic} - \textit{Memory depth}: \(\infty\). \cite{Prison1998}
\item Prober 4 - \textit{Deterministic} - \textit{Memory depth}: \(\infty\). \cite{Prison1998}
\item Pun1 - \textit{Deterministic} - \textit{Memory depth}: 2. \cite{Ashlock2006}
\item Punisher - \textit{Deterministic} - \textit{Memory depth}: \(\infty\). \cite{axelrodproject}
\item Raider - \textit{Deterministic} - \textit{Memory depth}: 3. \cite{Ashlock2014}
\item Random Hunter - \textit{Deterministic} - \textit{Memory depth}: \(\infty\). \cite{axelrodproject}
\item Random: 0.1 - \textit{Stochastic} - \textit{Memory depth}: 0.
\item Random: 0.3 - \textit{Stochastic} - \textit{Memory depth}: 0.
\item Random: 0.5 - \textit{Stochastic} - \textit{Memory depth}: 0. \cite{Axelrod1980, Tzafestas2000}
\item Random: 0.7 - \textit{Stochastic} - \textit{Memory depth}: 0.
\item Random: 0.9 - \textit{Stochastic} - \textit{Memory depth}: 0.
\item Remorseful Prober: 0.1 - \textit{Stochastic} - \textit{Memory depth}: 2. \cite{Li2011}
\item Resurrection - \textit{Deterministic} - \textit{Memory depth}: 5. \cite{Eckhart2015}
\item Retaliate 2: 0.08 - \textit{Deterministic} - \textit{Memory depth}: \(\infty\). \cite{axelrodproject}
\item Retaliate 3: 0.05 - \textit{Deterministic} - \textit{Memory depth}: \(\infty\). \cite{axelrodproject}
\item Retaliate: 0.1 - \textit{Deterministic} - \textit{Memory depth}: \(\infty\). \cite{axelrodproject}
\item Revised Downing: True - \textit{Deterministic} - \textit{Memory depth}: \(\infty\). \cite{Axelrod1980}
\item Ripoff - \textit{Deterministic} - \textit{Memory depth}: 2. \cite{Ashlock2008}
\item Risky QLearner - \textit{Stochastic} - \textit{Memory depth}: \(\infty\). \cite{axelrodproject}
\item SelfSteem - \textit{Stochastic} - \textit{Memory depth}: \(\infty\). \cite{Andre2013}
\item ShortMem - \textit{Deterministic} - \textit{Memory depth}: 10. \cite{Andre2013}
\item Shubik - \textit{Deterministic} - \textit{Memory depth}: \(\infty\). \cite{Axelrod1980}
\item Slow Tit For Two Tats - \textit{Deterministic} - \textit{Memory depth}: 2. \cite{axelrodproject}
\item Slow Tit For Two Tats 2 - \textit{Deterministic} - \textit{Memory depth}: 2. \cite{Prison1998}
\item Sneaky Tit For Tat - \textit{Deterministic} - \textit{Memory depth}: \(\infty\). \cite{axelrodproject}
\item Soft Go By Majority - \textit{Deterministic} - \textit{Memory depth}: \(\infty\). \cite{Axelrod1984, Mittal2009}
\item Soft Go By Majority: 10 - \textit{Deterministic} - \textit{Memory depth}: 10. \cite{axelrodproject}
\item Soft Go By Majority: 20 - \textit{Deterministic} - \textit{Memory depth}: 20. \cite{axelrodproject}
\item Soft Go By Majority: 40 - \textit{Deterministic} - \textit{Memory depth}: 40. \cite{axelrodproject}
\item Soft Go By Majority: 5 - \textit{Deterministic} - \textit{Memory depth}: 5. \cite{axelrodproject}
\item Soft Grudger - \textit{Deterministic} - \textit{Memory depth}: 6. \cite{Li2011}
\item Soft Joss: 0.9 - \textit{Stochastic} - \textit{Memory depth}: 1. \cite{Prison1998}
\item SolutionB1 - \textit{Deterministic} - \textit{Memory depth}: 3. \cite{Ashlock2015}
\item SolutionB5 - \textit{Deterministic} - \textit{Memory depth}: 5. \cite{Ashlock2015}
\item Spiteful Tit For Tat - \textit{Deterministic} - \textit{Memory depth}: \(\infty\). \cite{Prison1998}
\item Stochastic Cooperator - \textit{Stochastic} - \textit{Memory depth}: 1. \cite{Adami2013}
\item Stochastic WSLS: 0.05 - \textit{Stochastic} - \textit{Memory depth}: 1. \cite{axelrodproject}
\item Suspicious Tit For Tat - \textit{Deterministic} - \textit{Memory depth}: 1. \cite{Beaufils1997, Hilbe2013}
\item TF1 - \textit{Deterministic} - \textit{Memory depth}: \(\infty\). \cite{axelrodproject}
\item TF2 - \textit{Deterministic} - \textit{Memory depth}: \(\infty\). \cite{axelrodproject}
\item TF3 - \textit{Deterministic} - \textit{Memory depth}: \(\infty\). \cite{axelrodproject}
\item Tester - \textit{Deterministic} - \textit{Memory depth}: \(\infty\). \cite{Axelrod1980b}
\item ThueMorse - \textit{Deterministic} - \textit{Memory depth}: \(\infty\). \cite{axelrodproject}
\item ThueMorseInverse - \textit{Deterministic} - \textit{Memory depth}: \(\infty\). \cite{axelrodproject}
\item Thumper - \textit{Deterministic} - \textit{Memory depth}: 2. \cite{Ashlock2008}
\item Tit For 2 Tats (\textbf{Tf2T}) - \textit{Deterministic} - \textit{Memory depth}: 2. \cite{Axelrod1984}
\item Tit For Tat (\textbf{TfT}) - \textit{Deterministic} - \textit{Memory depth}: 1. \cite{Axelrod1980}
\item Tricky Cooperator - \textit{Deterministic} - \textit{Memory depth}: 10. \cite{axelrodproject}
\item Tricky Defector - \textit{Deterministic} - \textit{Memory depth}: \(\infty\). \cite{axelrodproject}
\item Tullock: 11 - \textit{Stochastic} - \textit{Memory depth}: 11. \cite{Axelrod1980}
\item Two Tits For Tat (\textbf{2TfT}) - \textit{Deterministic} - \textit{Memory depth}: 2. \cite{Axelrod1984}
\item VeryBad - \textit{Deterministic} - \textit{Memory depth}: \(\infty\). \cite{Andre2013}
\item Willing - \textit{Stochastic} - \textit{Memory depth}: 1. \cite{Berg2015}
\item Win-Shift Lose-Stay: D (\textbf{WShLSt}) - \textit{Deterministic} - \textit{Memory depth}: 1. \cite{Li2011}
\item Win-Stay Lose-Shift: C (\textbf{WSLS}) - \textit{Deterministic} - \textit{Memory depth}: 1. \cite{Kraines1989, Nowak1993, Stewart2012}
\item Winner12 - \textit{Deterministic} - \textit{Memory depth}: 2. \cite{Mathieu2015}
\item Winner21 - \textit{Deterministic} - \textit{Memory depth}: 2. \cite{Mathieu2015}
\item Worse and Worse - \textit{Stochastic} - \textit{Memory depth}: \(\infty\). \cite{Prison1998}
\item Worse and Worse 2 - \textit{Stochastic} - \textit{Memory depth}: \(\infty\). \cite{Prison1998}
\item Worse and Worse 3 - \textit{Stochastic} - \textit{Memory depth}: \(\infty\). \cite{Prison1998}
\item ZD-Extort-2 v2: 0.125, 0.5, 1 - \textit{Stochastic} - \textit{Memory depth}: 1. \cite{Kuhn2017}
\item ZD-Extort-2: 0.1111111111111111, 0.5 - \textit{Stochastic} - \textit{Memory depth}: 1. \cite{Stewart2012}
\item ZD-Extort-4: 0.23529411764705882, 0.25, 1 - \textit{Stochastic} - \textit{Memory depth}: 1. \cite{axelrodproject}
\item ZD-GEN-2: 0.125, 0.5, 3 - \textit{Stochastic} - \textit{Memory depth}: 1. \cite{Kuhn2017}
\item ZD-GTFT-2: 0.25, 0.5 - \textit{Stochastic} - \textit{Memory depth}: 1. \cite{Stewart2012}
\item ZD-SET-2: 0.25, 0.0, 2 - \textit{Stochastic} - \textit{Memory depth}: 1. \cite{Kuhn2017}

	\end{enumerate}
\end{multicols}

\end{document}